# Investigation of Ultrafast Demagnetization and Gilbert Damping and their Correlation in Different Ferromagnetic Thin Films Grown UnderIdentical Conditions[1]

Suchetana Mukhopadhyay, Sudip Majumder, Surya Narayan Panda, Anjan Barman*

*Department of Condensed Matter and Materials Physics, S.N. Bose National Center for Basic Sciences, Block-JD, Sector III, Salt Lake, Kolkata 700106, India*
Email: abarman@bose.res.in

## Abstract

Following the demonstration of laser-induced ultrafast demagnetization in ferromagnetic nickel, several theoretical and phenomenological propositions have sought to uncover its underlying physics. In this work we revisit the three temperature model (3TM) and the microscopic three temperature model (M3TM) to perform a comparative analysis of ultrafast demagnetization in 20-nm-thick cobalt, nickel and permalloy thin films measured using an all-optical pump- probe technique. In addition to the ultrafast dynamics at the femtosecond timescales, the nanosecond magnetization precession and damping are recorded at various pump excitation fluences revealing a fluence-dependent enhancement in both the demagnetization times and the damping factors. We confirm that the Curie temperature to magnetic moment ratio of a given system acts as a figure of merit for the demagnetization time, while the demagnetization times and damping factors show an apparent sensitivity to the density of states at the Fermi level for a given system. Further, from numerical simulations of the ultrafast demagnetization based on both the 3TM and the M3TM, we extract the reservoir coupling parameters that best reproduce the experimental data and estimate the value of the spin flip scattering probability for each system. We discuss how the fluence-dependence of inter-reservoir coupling parameters so extracted may reflect a role played by nonthermal electrons in the magnetization dynamics at low laser fluences.

## 1. Introduction

Optical excitation of a magnetic material with a short and intense laser pulse sets in motion a chain of microscopic events that result in a macroscopic, measurable quenching of the net magnetization. Since the processing speed of magnetic storage devices is limited by the maximum speeds at which the magnetization can be manipulated, the possibility of controlling magnetization at sub-picosecond timescales by employing ultrashort laser pulses holds tremendous potential for applications in spin-based memory and storage devices with ultrafast processing speeds including laser-induced opto-magnetism and all-optical switching [1–4] as well as THz spintronic devices [5–7]. Meanwhile, the relaxation timescales of the laser-induced magnetization precession in ferromagnetic thin films associated with the magnetic damping set fundamental limits on magnetization switching and data transfer rates in spintronic devices. Picosecond laser-induced magnetization quenching in ferromagnetic gadolinium was reported by Vaterlaus et al. in 1991, who also reported for the first time a characteristic timescale of 100±80 ps [8] associated with the magnetization loss. In this early work, the timescale was identified with the timescale of the spin-lattice interactions mediating the disruption of magnetic ordering due to laser heating of the lattice, though later works contradicted this interpretation [9, 10]. In fact, by the mid-1990s, it had begun to be recognized that finer time resolution was necessary to probe this phenomenon to uncover a fuller picture of the associated relaxation processes occurring at sub-picosecond timescales. In 1996, Beaurepaire et al. demonstrated in a seminal work that faster demagnetization occurring at sub-picosecond timescales could be triggered by femtosecond laser pulses in a nickel thin film [11]. The ultrafast demagnetization phenomenon went on to be demonstrated in a wide array of ferromagnetic systems [12–19], triggering a flurry of research that continues till date. A few years after the pioneering experiments, Koopmans et al. characterized for the first time the full magneto-optical response to femtosecond laser pulses in a ferromagnet by time-resolved measurements of Kerr ellipticity and rotation [13]. However, the observation of nonmagnetic contributions to the Kerr rotation signal naturally led some to question whether the ultrafast quenching in response to the laser excitation was indeed a magnetic phenomenon [13, 14]. In 2003, Rhie et al. used time-resolved photoemission spectroscopy to probe the collapse of the 3d exchange splitting in nickel as an unambiguous signature of a photoinduced demagnetization occurring over 300±70 fs [15]. This was soon followed by the first reports of an estimate of the characteristic timescale of femtosecond laser-induced demagnetization derived from quantum-mechanical principles [16].

---

[1] First submitted in March 2022

One of the major reasons behind this sustained interest is the need to achieve a complete understanding of the microscopic mechanisms that underlie the ultrafast demagnetization phenomenon that have so far remained elusive. Much inquiry has been focused on how the laser-induced loss of magnetic order is compensated for via the transfer of angular momentum of the spin system at the associated timescales. Over the years, several mechanisms have been proposed for explaining the angular momentum conservation associated with the ultrafast demagnetization, which may be broadly categorized in two distinct classes. One of these argues in favor of a dominant contribution to the demagnetization arising from nonlocal transport processes driven by the laser pulse, such as superdiffusive transport of spin-polarized hot electrons [20, 21] or heat currents [22]. The other narrative relies on local spin-flip scattering processes occurring by the collision of excited electrons with impurities, phonons, and magnons [17, 23, 24]. In this category, electron-lattice and electron-spin relaxation mechanisms are accepted as the major demagnetization channels in the picoseconds following the laser excitation [25], with several works attributing a pivotal role to the material-specific spin-orbit interaction [23, 25–27]. On the other hand, it is possible to model the demagnetization by adopting a thermal description, considering the laser excitation as a "heat" source that excites electrons close to the Fermi level instantaneously to very high energies, whose eventual relaxation processes drive the magnetization loss. The laser excitation energy controlled by the applied laser pump fluence has a direct influence on the maximum electron temperatures reached and thus plays a pivotal role in the ultrafast magnetization dynamics that follow as a result.

In this work, employing an all-optical time-resolved magneto-optical Kerr effect (TR-MOKE) technique, we investigate the ultrafast spin dynamics along with the nanosecond magnetization precession and damping at different laser pump fluences in three ferromagnetic thin films: 20-nm-thick cobalt, nickel and permalloy grown under uniform deposition conditions. Cobalt and nickel are elementary 3d ferromagnetic transition metals well studied in the literature in both their elementary forms and as constituting elements in multilayered structures where their strong spin-orbit coupling mediates interfacial effects such as spin pumping. Permalloy, an alloy comprised of approximately 80% nickel and 20% iron, is a prototype material for spintronic applications due to several desirable qualities such as low coercivity, large permeability and in particular, low Gilbert damping, which aids in minimizing the maximum power consumption of devices and allows spin wave propagation on length scales of the order of device size. The simultaneous investigation of the fluence-dependent modulation of ultrafast demagnetization and damping in all the above ferromagnetic thin films is motivated by the primary objectives of exploring the dominant microscopic processes underlying the demagnetization in these systems, correlating the observed magnetization dynamics to the material properties of each system, and in the process exploring their tunability with laser pump fluence. We set out to achieve this in two ways: (a) by modelling the ultrafast demagnetization at various pump fluences using two theoretical models sharing a common link and (b) by correlating the laser-induced changes in ultrafast demagnetization to the changes in the Gilbert damping factor which characterizes the magnetization precession. TR-MOKE is a well established local and non-invasive all-optical measurement technique tunable to an extremely fine time resolution limited only by the laser pulse width and is thus well suited to our purpose.

For the detailed analysis of ultrafast demagnetization in our samples, we extract the values of demagnetization time and fast relaxation time for each sample using a phenomenological fitting function and record its systematic variation with the pump fluence. We subsequently model the demagnetization data using two well known theoretical models: the phenomenological three-temperature model (3TM) and the microscopic three temperature model (M3TM), and thereby extract values for the microscopic parameters relevant for the demagnetization process and calculate the temporal evolution of the simulated temperatures of the electron, spin and lattice systems within the first few picoseconds of the laser excitation. Both these models assume a thermal picture to explain the initial electron temperature rise and subsequent relaxation but differ significantly in their approach with regards to the treatment of the magnetization. A systematic study implementing both models to analyze TR-MOKE data recorded under uniform experimental conditions on identically prepared samples is absent in the literature and will prove instructive. Further, a thorough investigation and characterization of the ultrafast demagnetization, magnetic damping and their intercorrelation in three different systems deposited under identical deposition conditions has not been carried out before. For the purposes of a comparative analysis, it is vital to study samples grown under the same deposition conditions. The conductivity of the substrate can also directly influence demagnetization time by promoting or hindering ultrafast spin transport processes [20, 28] while the deposition technique determines the structural properties of the samples that may indirectly affect the demagnetization time. Moreover, since the experimentally obtained demagnetization timescale has been shown to be quite sensitive to extrinsic factors such as the laser pulse duration and the spectral bandwidth [29], it is useful to measure the demagnetization times for various thin films from experiments performed in the same experimental setup under near-identical experimental conditions.

## 2. Materials and Methods

*2.1 Sample fabrication*

20 nm-thick cobalt, nickel and permalloy thin films were deposited by electron beam evaporation under an average base pressure of $1\times10^{-7}$ Torr and at a very low deposition rate of 0.2 Å/s chosen to achieve uniform deposition. Each film was deposited on an insulating 8 mm × 8 mm silicon substrate coated with 285-nm-thick $SiO_2$. Subsequently, the films were capped *in-situ* with a 5-nm-thick protective gold layer (base pressure ~$1 \times 10^{-7}$ Torr, deposition rate 0.1 Å/s) to prevent surface oxidation of the ferromagnetic layer and protect it from possible degradation due to high power laser exposure during the optical pump-probe measurements [30, 31]. The thickness of the capping layer was kept more than three times smaller than the optical penetration depth of 400 nm light in gold. After the deposition of the capping layer, the surface topography of the samples was investigated by atomic force microscopy (AFM) using a Tap190Al-G tapping mode AFM probe as shown in Figure 1 (*a*). The average surface roughness $R_a$ of the cobalt, nickel and permalloy films were obtained as 0.91 nm, 0.36 nm and 0.41 nm respectively. Thus, reasonably low surface roughness in the sub-nm range was obtained which is comparable for all the samples. In addition, the static magneto-optical Kerr effect was used to study the magnetic hysteresis of the deposited samples to confirm their ferromagnetic nature (Figure 1(*b*)).

*2.2 All-optical measurement of ultrafast spin dynamics*

Measurement of laser-induced magnetization dynamics is carried out using a TR-MOKE technique in two-color optical pump-probe arrangement using a 400 nm pump beam and 800 nm probe beam having a pump-probe cross-correlation width of ~100 fs [30] shown schematically in Figure 1(*c*). A typical TR-MOKE trace is shown in Figure 1(*d*), comprising ultrafast demagnetization (Regime I), fast remagnetization (Regime II) and damped magnetization precession (Regime III).This technique enables the direct observation of the spin dynamics in the femto- and picosecond time domain. Femtosecond pulses are generated by an amplified laser system (Libra, Coherent Inc.) employing a chirped-pulse regenerative amplifier and a Ti:sapphire laser oscillator (Coherent Inc.) pumped by a neodymium-doped yttrium lithium fluoride (Nd-YLF) laser. The second harmonic of the amplified laser output (wavelength = 400 nm, repetition rate = 1 kHz, pulse width >40 fs), generated through a lithium triborate (LBO) nonlinear crystal, is used for laser excitation of the ferromagnetic thin films. The time-delayed fundamental beam (wavelength = 800 nm, repetition rate = 1 kHz, pulse width ~40 fs) is used to probe the ensuing magnetization dynamics. In our setup, different wavelengths are employed for the pump and the probe pulse to eliminate the possibility of state blocking effects arising from the use of identical wavelengths for pumping and probing [32]. A computer-controlled variable delay generator offers precise control of the delay time between pump and probe. Before commencing measurements on any sample, the zero delay was carefully estimated by maximizing the transient reflectivity signal of a bare silicon substrate placed adjacent to the sample on the same sample holder. TR-MOKE experiments are performed with a non-collinear pump-probe geometry. The pump beam, focused to a spot size of ~300 μm, is incident obliquely on the sample while the probe beam, with a spot size of ~100 μm, is incident normal to the sample surface and aligned to the center of the pump spot. The pump-probe spatial overlap on the sample was carefully maintained. The choice of a relatively smaller spot size of the probe beam as compared to the pump beam facilitates optical alignment and ensures that the probe beam detects the local magnetization changes from a part of the sample uniformly irradiated by the pump. Before reflection on the samples, the probe beam is polarized orthogonally to the linearly polarized pump beam. After reflection, the Kerr-rotated probe beam is split and one part is fed directly into a Si-photodiode to measure the time-resolved reflectivity signal. The other part is fed into an identical photodiode after passing through a Glan-Thompson polarizer adjusted to a small angle from extinction to minimize optical artifacts in the Kerr rotation signal. In this way, simultaneous measurement of the time-resolved total reflectivity and Kerr rotation signals is possible. An optical chopper operating at 373 Hz placed in the path of the pump beam provides a reference signal for the digital signal processing lock-in amplifiers (Stanford Research Systems, SR830) which record the modulated signal in a phase sensitive manner. All experiments were carried out under ambient conditions of temperature and pressure.

## 3. Results and Discussion

### 3.1 Theoretical models for ultrafast demagnetization

The phenomenon of optically induced ultrafast demagnetization starts with the irradiation of the magnetic sample with a brief and intense optical laser pulse, exciting electrons momentarily a few electron volts above the Fermi level. Though the exact sequence of events following the initial excitation is difficult to trace due to the highly nonequilibrium conditions created by it, a qualitative overview of the complete demagnetization process is fairly well established. The laser excitation generates a nonthermal pool of excited electrons which thermalize rapidly within several femtoseconds via electron-electron interactions. Spin-dependent scattering events taking place during this transient regime lead to a sharp drop in magnetization observable around a few hundred femtoseconds in the experimental Kerr rotation signal. Subsequently, the thermalized electrons may release their excess energy via a variety of relaxation channels, such as by excitation of phonons or magnons. This results in a partial recovery of the magnetization beyond which heat dissipation into the environment promotes further recovery on a longer timescale. The 3TM posits that the thermodynamics of the demagnetization phenomenon can be described simply by considering energy exchange between three thermal reservoirs [33], each of which is assigned a temperature: the electrons at temperature $T_e$, the lattice at temperature $T_l$ and the electronic spin reservoir at temperature $T_s$. Since the reservoirs are in thermal contact and the overall process is adiabatic, equilibration of the excited electrons with the spin and lattice reservoirs via energy transfer may be described by coupled rate equations in the following manner:

$$C_e(T_e)\frac{dT_e}{dt} = -G_{el}(T_e - T_l) - G_{es}(T_e - T_s) + \nabla_z(\kappa \nabla_z T_e) + P(t)$$

$$C_l(T_l)\frac{dT_l}{dt} = -G_{el}(T_l - T_e) - G_{ls}(T_l - T_s) \qquad (1)$$

$$C_s(T_s)\frac{dT_s}{dt} = -G_{es}(T_s - T_e) - G_{ls}(T_s - T_l)$$

where, $C_e$, $C_l$ and $C_s$ denote the specific heats of the electron, lattice and spin reservoirs respectively, while $G_{el}$, $G_{es}$, and $G_{sl}$ denote the inter-reservoir coupling parameters. The term $P(t)$ describes the action of the laser pulse as a source term driving the excitation of the electron reservoir to high temperatures. The thermal diffusion term $\nabla_z(\kappa \nabla_z T_e)$ describes heat dissipation occurring via thermal conduction along the sample thickness. Under this description, the observed demagnetization is attributed to a rise in the spin temperature $T_s$ occurring shortly after the electron temperature rise. The coupling strengths between the electron-spin and electron-lattice subsystems qualitatively determine the efficiency of energy transfer between them and hence influence the timescales associated with the demagnetization and fast relaxation. However, the 3TM is purely phenomenological and does not explicitly consider any microscopic mechanisms underlying the phenomenon it describes. On the other hand, the M3TM proposed by Koopmans et al. [23] provides a "spin-projected" perspective [27] to explain the ultrafast transfer of angular momentum highlighting the role of Elliot-Yafet type ultrafast spin-flip scattering in the demagnetization process. The initial excitation by the laser pulse disturbs the electronic subsystem from equilibrium which leads to an imbalance in spin-up and spin-down scattering rates, resulting in the observed loss of magnetic order. The process is mediated by spin-orbit interactions leading to the formation of hot spots in the band structure where spin-up and spin-down channels are intermixed. An electron scattered into these hot spots via a phonon- or impurity-mediated scattering will flip its spin with a finite probability. The individual scattering events are characterized by a parameter $a_{sf}$ across the sample, identified with the probability of spin-flip due to electron-phonon scattering. The magnitude of this parameter will directly depend on the extent of spin-orbit coupling and hence is expected to be comparable in materials with similar spin-orbit coupling strengths. The M3TM retains the coupled rate equations for the electron and lattice temperatures, similar to the thermal description provided by the 3TM. However the fundamental difference from the 3TM is that in the framework of the M3TM, the spin bath is formed by a collection of two-level systems obeying Boltzmann statistics. Instead of assigning a temperature to the spin bath, the normalized magnetization is directly calculated from the associated exchange splitting. The rate of change of the magnetization, derived analytically considering an Elliot-Yafet scattering-driven demagnetization, is parametrized by $a_{sf}$ and coupled to the electron and the lattice subsystem temperatures. The assignment of a characteristic temperature to the spin subsystem is replaced in the M3TM by an evolution equation for the magnetization:

$$\frac{dm}{dt} = Rm\frac{T_l}{T_C}\left(1 - \coth\left(\frac{mT_C}{T_e}\right)\right) \tag{2}$$

The quantity *R* is a material-specific factor which influences the demagnetization rate and is proportional to $a_{sf}T_c^2/\mu_{at}$ where $T_C$ is the material Curie temperature and $\mu_{at}$ is the atomic magnetic moment. Though the two models differ in their approaches, one can immediately discern certain similarities in their domains of validity. Both models are appropriate only when the nonlocal mechanisms driving ultrafast demagnetization such as superdiffusive spin transport can be neglected. We note here that it has been reported that spin transport is not a major contributor to the ultrafast demagnetization in transition metals [34]. We nevertheless use an insulating $SiO_2$-coated Si substrate for our samples to minimize spin transport effects such that analysis with the local models described above should suffice in our case. Any additional contributions arising from the gold capping layer would be uniform across all samples investigated and therefore unlikely to impact the main results of our comparative study. Moreover, since the thickness of the capping layer is much smaller than the penetration depth of both 400 nm and 800 nm light in gold [35], the pump excitation fully penetrates down to the magnetic layer ensuring that the effect of direct laser excitation of the ferromagnet is probed in our case. Thus, we set up numerical calculations based on the models described above in order to extract microscopic information from the experimentally obtained demagnetization traces.

*3.2 Ultrafast demagnetization in cobalt, nickel, and permalloy thin films*

We proceed by performing time-resolved measurements of the polar magneto-optical Kerr effect in the cobalt, nickel and permalloy thin film samples as a function of the laser fluence. Measurements are carried out under a strong enough external magnetic field kept constant at around 2 kOe tilted at a small angle from the sample plane to saturate the magnetization of the samples. The pump fluence is varied between 0.8-8.7 mJ/cm² by varying the power of the pump pulse. The results are presented in Figure 2. To ascertain that the measured Kerr signal reflects the true magnetization dynamics without any spurious contribution from optical effects triggered in the initial stages of laser excitation, we also examine the transient reflectivity signal for each sample. The fluence dependent variation in the reflectivity can be found in Figure S1 of the Supplementary Materials, demonstrating that at any given fluence the amplitude of the reflectivity signal is negligibly small compared to that of the Kerr rotation. We nevertheless restrict ourselves to the low fluence regime to avoid nonlinear effects and sample damage. For all our experiments, the probe fluence is kept constant at a value about half that of the lowest pump fluence used to prevent additional contribution to the spin dynamics by probe excitation. As seen in Figure 2, the ultrafast demagnetization completes within 1 ps for all three samples considered which is followed by a fast recovery of the magnetization, all observed within the experimental time window of 4 ps. These experimental traces clearly exhibit the "Type-I" or "one-step" demagnetization expected for transition metal thin films at room temperature and under low-to-moderate pump fluence [23]. The amplitude of the maximum quenching of the Kerr rotation signal increases with the laser fluence, allowing us to rule out nonlinear effects [36]. Closer inspection of the traces also reveals an increase of the time taken to demagnetize the samples with increasing fluence for all three samples. To quantify this increase, we fit our demagnetization traces to a phenomenological expression based on the 3TM and valid in the low laser fluence regime [37]:

$$-\frac{\Delta M_z}{M_z} = \left[\left\{\frac{A_1}{(t/\tau_0 + 1)^{1/2}} - \frac{(A_2\tau_E - A_1\tau_M)e^{-t/\tau_M}}{\tau_E - \tau_M} - \frac{\tau_E(A_1 - A_2)e^{-t/\tau_E}}{\tau_E - \tau_M}\right\}\Theta(t)\right] \otimes \Gamma(t) \tag{3}$$

where $\Theta(t)$ is the Heaviside step function, $\delta(t)$ is the Dirac delta function and $\Gamma(t)$ is the Gaussian laser pulse. The constant $A_1$ represents the value of the normalized magnetization after remagnetization has completed and equilibrium between the electron, spin and lattice reservoirs has been re-established. $A_2$ is proportional to the initial rise in electron temperature and hence to the maximum magnetization quenching. $A_3$ represents the magnitude of state-filling effects present during the onset of the demagnetization response, which is negligible in our case. $\tau_M$ and $\tau_E$ are the demagnetization time and fast relaxation time, respectively. Prior to the fitting, all the experimental traces were normalized by hysteresis measurements of the Kerr rotation signal under the saturating magnetic field in the absence of laser excitation. We find that within the range of fluence values considered, permalloy exhibits the largest magnetization quenching of 54.6%, followed by a 23.7% quenching achieved in nickel, while the magnetization of cobalt the least, only about 8% for the largest applied fluence. The demagnetization occurs at a characteristic timescale of 230-280 fs for cobalt, 160-210 fs for nickel, and 220-250 fs for permalloy, increasing with the laser fluence. This effect can be attributed to enhanced spin-fluctuations at elevated spin temperatures for higher fluences [38]. At a fluence of 4.8 mJ/cm$^2$, the extracted demagnetization times are 276.6 ± 3.41 fs for cobalt, 187.3 ± 2.89 fs for nickel and 236.8 ± 2.45 fs for permalloy. The timescale for the magnetization recovery $\tau_E$ also increases with increasing pump fluence. The variation of these characteristic timescales with laser fluence is shown in Figure 3. These fluence-dependent trends in $\tau_M$ and $\tau_E$ hint at a spin-flip process-dominated ultrafast demagnetization in our studied systems [23, 39, 40]. The values of $\tau_M$ extracted from our experiments lie within the typical range of 100-300 fs consistent with previous reports of the ultrafast demagnetization times in these metals [17, 23], and are too large to represent a superdiffusive transport-driven demagnetization [41].

For the 3TM and M3TM simulations, we choose a laser pump term given by $P(t) = P(0)Fe^{-(t/\tau_p)^2}$ proportional to the pump fluence $F$ and following a Gaussian temporal profile. The maximum rise of the electron temperature and thus also the extent of demagnetization depends sensitively on this term which is hence adjusted to reproduce the maximum quenching observed experimentally. We use a pulse width $\tau_p = 100$ fs determined by the pump-probe cross-correlation in all calculations. Intrinsic to both the models we consider is the assumption that electron thermalization occurs extremely fast. The thermal diffusion term can be neglected in our case since the thicknesses of the films we study are kept slightly greater than the optical penetration depth of 400 nm pump beam in those films. This ensures uniform heating of the films in the vertical direction while also avoiding laser penetration into the substrate in which case heat dissipation into the substrate would have to be taken into account. Besides, the timescales associated with heat dissipation are generally tens to hundreds of picoseconds, much longer than the demagnetization and fast relaxation times, and hence unlikely to significantly influence our observations at these timescales. Since both models we consider are thermal in their approach, choosing correct values for the reservoir specific heats is vital for a proper simulation of the demagnetization. For the electronic specific heat $C_e$, we assume a linear dependence on the electronic temperature $C_e(T_e) = \gamma T_e$ derived from the Sommerfeld free-electron approximation where $\gamma$ is determined by the electronic density of states at the Fermi level [42]. The value of $\gamma$ for permalloy is approximated as a weighted average of the individual $\gamma$ values of nickel and iron in permalloy. The lattice specific heat $C_l$ is calculated at each value of the lattice temperature according to the following relation derived from Debye theory:

$$C_l(T_l) = 9N_A k_B \left(\frac{T_l}{\theta_D}\right)^3 \int_0^{\frac{T_l}{\theta_D}} \frac{x^4 e^x}{(e^x - 1)^2} dx \qquad (4)$$

where $N_A$ is the Avogadro's number, $k_B$ is the Boltzmann's constant and $\theta_D$ is the Debye temperature. Finally, we fix the spin specific heat $C_s$ to its value at room temperature for the 3TM calculations, obtained by subtracting the electronic and lattice contributions from the experimental values of the total specific heat found in the literature [43]. Considering a spin-temperature-dependent form of $C_s$ was not found to significantly affect our conclusions as described in Section IV of the Supplementary Materials. The fixed parameter set used in our calculations have been listed in Table **1**. To relate the experimental demagnetization to the temperature of the spin subsystem under the 3TM framework, the spin temperature $T_s$ is mapped to the magnetization of the system via the Weiss' mean field theory [44], which is then fitted with the experimental magnetization traces to obtain the empirical inter-reservoir coupling parameters $G_{el}$ and $G_{es}$ consistent with the observed dynamics. We neglect the spin-lattice coupling parameter $G_{sl}$ for the 3TM simulations since in ferromagnetic transition metals the energy transfer between electrons and lattice is far greater than that between lattice and spins [45].

**Table 1**. Fixed parameter set used in the calculations. Literature values have been used for all parameters listed [30, 31, 34, 35].

| Sample | $T_c$ (K) | $\theta_D$ (K) | $\gamma$ (Jm$^{-3}$K$^{-2}$) | $C_s$ (Jm$^{-3}$K$^{-1}$) | $\mu_{at}$ ($\mu_B$) |
|---|---|---|---|---|---|
| Ni | 627 | 450 | 1065 | 3.07 × 10$^5$ | 0.62 |
| Co | 1388 | 445 | 714 | 1.59 × 10$^5$ | 1.72 |
| Py | 860 | 454 | 992 | 2.67 × 10$^5$ | 1.00 |

For the 3TM simulations, we proceeded to extract $G_{el}$ and $G_{es}$ by first fitting the demagnetization data at the lowest fluence to the model. However, fitting the higher fluence data using identical values of the coupling parameters as extracted at the lowest fluence did not result in a good match to our experimental results. The coupling parameters extracted from the low fluence data led to an overestimation of the demagnetization time at the higher fluences. It was seen that a 5-10% increase in $G_{es}$ from its value at its adjacent lower fluence value rectifies the overestimation of the demagnetization time. On the other hand, the remagnetization dynamics is most sensitive to the $G_{el}$ parameter so that the overall dynamics is best reproduced only by adjusting both $G_{el}$ and $G_{es}$. As shown in Figure 4, the resulting fit shows excellent agreement with the experimental data. This exercise reveals the crucial role played by the electron-spin relaxation channels in determining the timescale associated with the initial demagnetization while the magnetization recovery is primarily mediated by the electron-lattice interaction. We also find that the mismatch between model and experiment can be resolved by considering an increasing trend of $G_{el}$ and $G_{es}$ with pump fluence arising from a faster demagnetization process for the same percentage quenching as compared to the model predictions within the studied fluence range. The values of the microscopic parameters extracted from the least-squares fits with their corresponding error bounds can be found in Supplementary Tables S1-S3. Since the exact values of the coupling parameters extracted from the fits naturally depend on the values chosen for the fixed parameters, the interpretation of the results from these fits is best limited to a comparative one.

For the M3TM simulations, the demagnetization traces are fitted directly to Equation 2 yielding $G_{el}$ and $a_{sf}$ as fit parameters. In this case, $a_{sf}$ plays a role in determining the maximum extent and the associated timescale of the demagnetization via the scaling factor $R$ while $G_{el}$ continues to influence mainly the magnetization recovery process. However, the demagnetization time is less sensitive to changes in $a_{sf}$ than it is to $G_{es}$ in the 3TM case. This results in somewhat higher values of $G_{el}$ and a sharper rise with pump fluence than those extracted from the 3TM simulations, in order to compensate for the overestimation of demagnetization time that results from the model if $G_{el}$ at the lowest fluence is used for all the fits. We have obtained an $a_{sf}$ of ~0.02 for cobalt, ~0.05-0.06 for nickel, and ~0.03-0.06 for permalloy. The value of $a_{sf}$ we have extracted for nickel is an order of magnitude lower than the value $a_{sf} = 0.185$ first reported by Koopmans et al. [23] but quite close to the value of 0.08 reported by Roth et al. [39]. This discrepancy is expected, as the artificially high value of 0.185 arose due to an overestimation of the electronic specific heat in the original work, avoided here by considering experimentally determined $\gamma$ values reported in the literature. The observation that $a_{sf}$[Co] $< a_{sf}$[Ni] is consistent with previous reports [23, 40].

With regard to thermal treatments of ultrafast demagnetization like the 3TM and M3TM, there is some contention centered on the role of the nonthermal electrons generated within the first 50-100 femtoseconds after laser excitation [9, 48, 49]. Within this time frame, the excited hot electrons have not yet thermalized and their energies cannot be described by a Fermi-Dirac distribution. However, both the 3TM and the M3TM completely neglect the finite electron thermalization time and consider only the energy exchange between the thermalized electron population with the lattice and/or the spin subsystem as the initial nonequilibrium distribution is assumed to last for much shorter durations ($\gtrsim$ 10 fs) than the timescale over which demagnetization typically occurs ($\gtrsim$ 100 fs). However, strictly speaking, this assumption is valid only when the thermalization proceeds very rapidly such as can be ensured with high fluence excitations [50]. In the low fluence regime, electron thermalization is slower, hence the influence of nonthermal electrons on the measured dynamics may become more important with decreasing pump fluence. The increasing of $G_{el}$ with fluence deduced from our calculations may be an indication of a decrease in the efficiency of electron-lattice interactions at low laser fluences when nonthermal electrons prevail. However, the specific values of $G_{el}$ and $G_{es}$ values we have obtained for the different systems remain comparable to values previously reported for these metals [11, 23, 37].

After completion of the fitting routine, we set up simulations using the optimized values of the free parameters to obtain the temporal evolution of $T_e$, $T_l$ and $T_s$ as a function of the pump-probe delay. The calculated profiles of $T_e$, $T_l$ and $T_s$ for four different values of the pump fluence are presented in Figure 5 for nickel, showing good agreement between the results predicted by the 3TM and the M3TM. Similar profiles for cobalt and permalloy are given in Figs. S2 and S3 of the Supplementary Materials. As the pump fluence is increased, laser excitation deposits a greater amount of energy in the electronic reservoir which leads to a proportionate increase in the initial electron temperature rise. However, the maximum electron temperature reached for the highest applied pump fluence remains well below the Fermi temperature for the respective material, ensuring the assumption of the linear temperature dependence of the electronic specific heat remains valid [42]. Given the fundamentally different origins of the 3TM and M3TM, the close agreement between the temperature profiles calculated using the two models and among the extracted values of the electron-lattice coupling parameter $G_{el}$ is remarkable. This observation reflects that, given the strength of the laser excitation determining the initial temperature excursion of the electron system, the $G_{el}$ parameter common to both the 3TM and the M3TM framework plays a pivotal role in determining the profile of the ensuing magnetization changes.

This also explains why $G_{el}$ is seen to follow a similar fluence-dependent trend for a given system in both models. While the electron temperature is coupled to the spin temperature via the $G_{es}$ parameter in the 3TM, it enters into the evolution equation for the magnetization via Equation 2 in the M3TM case. From Figure 5, a qualitative difference in the temporal profiles of the electron, spin and lattice temperatures is also apparent. Referring to the figures corresponding to the 3TM, while $T_e$ exhibits a sharp peak within the first few hundred femtoseconds, $T_s$ exhibits a shorter and broader peak at slightly longer time delays. The delayed peak in $T_s$ occurs when excited electrons transfer energy to the spin degrees of freedom [51] through scattering events mediated by the coupling parameter $G_{es}$, which qualitatively determines the maximum spin temperature value [11]. These scattering events can result in the generation of magnons, Stoner excitations and spin fluctuations which may play a part in the observed demagnetization [52]. Lastly, the lattice temperature gradually increases over the first few picoseconds after excitation, reaching an equilibrium value at which it stays nearly constant over several picoseconds.

In Figure 6 we present a comparison between the experimental demagnetization curves and calculated electron temperature profiles for each of our three samples. Under the same applied pump fluence, different amplitudes of magnetization quenching and electron temperature rise are observed. It is clearly seen that nickel quenches more strongly than cobalt, an observation which can be explained by band structure effects and a more efficient laser-induced electronic excitation in nickel owing to its much lower Curie temperature [34, 40, 53], reflected in the M3TM as a higher extracted value of the spin-flip probability $a_{sf}$ in nickel. On the other hand, a lower electronic specific heat capacity in cobalt as compared to nickel [42] predictably leads to a much greater rise in its electron temperature for the same pump fluence. In the case of permalloy, alloying with iron can reduce the electronic density of states at the Fermi level as compared to nickel, decreasing $\gamma$. Because of the larger electron temperatures reached as a result, a value of $a_{sf}$ comparable to that of nickel is sufficient to drive a much larger reduction in the magnetization. Permalloy also shows a somewhat larger demagnetization time at a given pump fluence as compared to nickel. Since a standard TR-MOKE optical pump-probe measurement cannot be used to probe the demagnetization response in alloys in an element-specific manner, interpreting this finding is difficult. It has been reported previously using element-specific measurements in a nickel-iron alloy that the demagnetization of nickel proceeds faster than that of iron [54]. Our observation of a slower demagnetization in permalloy may be reflective of this result. Finally, the demagnetization times for the three systems at any given pump fluence is seen to sensitively depend on the ratio of $T_C/\mu_{at}$, confirming its role as a 'figure of merit' (FOM) for the demagnetization time [23]. As shown in Figure 8 (*a*), the demagnetization proceeds slower for a lower value of the FOM. Thus at a given incident laser pump fluence, the demagnetization times are found to correlate with the values of this quantifying parameter, rather than the material Curie temperature directly.

*3.3 Precessional dynamics and correlation of Gilbert damping parameter with ultrafast demagnetization time*

To gain further insight into the microscopic mechanism underlying the ultrafast demagnetization, we study also the precessional magnetization dynamics in our thin film systems. A typical time-resolved Kerr rotation data showing the different temporal regimes of laser-induced magnetization dynamics is given in Figure 1(*d*). Regime I and II represent the ultrafast demagnetization and magnetization recovery processes respectively, while the oscillatory signal of regime III represents the magnetization precession with the Gilbert damping characteristic of the ferromagnetic material. To extract the Gilbert damping factor $\alpha$, the background-subtracted oscillatory Kerr signal is first fitted with a damped sinusoidal function given by:

$$M(t) = M(0) exp^{-(t/\tau)} \sin(2\pi f t + \varphi) \quad (5)$$

where *f* is the precessional frequency, $\tau$ is the relaxation time, and $\varphi$ is the initial phase of oscillation. The dependence of *f* on applied bias magnetic field *H* is fitted to the Kittel formula relevant for ferromagnetic systems:

$$f = \frac{\gamma_0}{2\pi} (H(H + 4\pi M_{eff}))^{1/2} \quad (6)$$

where $\gamma_0 = g\mu_B/\hbar$ is the gyromagnetic ratio and $M_{eff}$ is the effective saturation magnetization of the probed volume. From the Kittel fit, the values of *g* and $M_{eff}$ can be extracted as fit parameters. While the value of *g* is found to be $2 \pm 0.1$ for all the samples, the extracted values of $M_{eff}$ are 1280.2 ± 1.69 emu/cm$^3$ for cobalt, 400.5 ± 11.46 emu/cm$^3$ for nickel and 737.1 ± 5.32 emu/cm$^3$ for permalloy (Section V of the Supplementary Materials).

Subsequently, the effective damping factor $\alpha$ is calculated depending on the extracted value of $\tau$ and $M_{eff}$ as:

$$\alpha = \frac{1}{\gamma_0 \tau (H + 2\pi M_{eff})} \qquad (7)$$

Although the timescale of precessional dynamics and magnetic damping in ferromagnetic thinfilms differ by many orders of magnitude, similar underlying phenomena may often determine the characteristics of either process. Through several independent investigations relating these two distinct processes, the correlation between the demagnetization time $\tau_M$ and the Gilbert damping factor $\alpha$ emerged as a critical indicator of the dominant microscopic contribution to the ultrafast demagnetization. Koopmans et al. in 2005 analytically derived an inversely proportional relationship between the Gilbert damping factor and ultrafast demagnetization time based on quantum mechanical considerations [16]. However, the validity of the modelcould not be confirmed experimentally when $\tau_M$ and $\alpha$ were tuned by transition metal and rare earth doping [18]. Later it was proposed that the Gilbert damping parameter and ultrafast demagnetization time can be either directly or inversely proportional to each other depending on the predominant microscopic mechanism contributing to the magnetic damping [55]. A proportional dependence would suggest a dominating conductivity-like or "breathing Fermi surface" contribution to the damping due to the scattering of intraband electrons and holes, while an inverse dependence may arise in materials with a dominant resistivity-like damping arising from inter-band scattering processes. Zhang et al. later suggested that an inverse correlation between Gilbert damping factor and ultrafast demagnetization time may arise in the presence of spin transport, which provides an additional relaxation channel resulting in an enhancement of the magnetic damping along with an acceleration of the demagnetization process at the femtosecond timescale [56]. In cases where local spin-flip scattering processes are expected to dominate the demagnetization process, the breathing Fermi surface model is valid predicting a proportional relationship between $\tau_M$ and $\alpha$.

To study the correlation between ultrafast demagnetization and damping in a system, both $\tau_M$ and $\alpha$ have to be systematically varied. For a given material, the exciting pump fluence can be used to modulate the demagnetization time. As discussed in the previous section, we observe an increment of the demagnetization time with the pump fluence for the systems under investigation. An enhancement in $\alpha$ is also observed as the pump fluence is increased from 0.8 mJ/cm$^2$ to 8.7 mJ/cm$^2$ for each sample, associated with a corresponding decrease in the precessional relaxation time. Within the experimental fluence range, $\alpha$ is found to be enhanced from 0.0089 to 0.0099 for cobalt, from 0.0346 to 0.0379 for nickel and from 0.0106 to 0.0119 for permalloy. This enhancement of the damping can be attributed to the transient rise in the system temperature as the laser pump fluence is increased [57]. As a laser-induced increase in electron temperature also drives the demagnetization process at the femtosecond timescale, one can qualitatively correlate the demagnetization time with the magnetic damping at various excitation energies even for a single sample. Clearly, a direct correlation between demagnetization time and the Gilbert damping factor is obtained in our samples by measuring the magnetization dynamics at various pump fluences, as shown in Figure 7 for the nickel thin film. Thus, we infer that similar relaxation processes likely drive the laser induced transient magnetization dynamics at the two different timescales. At a low fluence value of 2.1 mJ/cm$^2$, the extracted value of $\alpha$ is the highest for the nickel film, and relatively low for cobalt and permalloy. The observed trend in Figure 8 (*b*) can be directly correlated with the values of the Sommerfeld coefficient $\gamma$ given in Table 1. This observation is in accordance with Kambersky's spin flip scattering theory [58], wherein $\alpha$ is directly proportional to the density of states at the Fermi level, which in turn governs the value of $\gamma$.

## 4. Summary

To summarize, we have investigated fluence-dependent ultrafast demagnetization in identically-grown cobalt, nickel and permalloy thin films using an all-optical time-resolved magneto-optical Kerr effect technique. We study both the femtosecond laser-induced ultrafast demagnetization and the magnetization precession and damping in each of our samples for various pump excitation fluences. Using a phenomenological expression, we extract the values of demagnetization time and fast relaxation time for each of our samples to show that they exhibit an increasing trend with applied laser fluence consistent with a spin-flip scattering-dominated demagnetization in our samples. Two distinct ultrafast demagnetization models can each reproduce our experimental demagnetization traces remarkably well. By directly fitting our experimental data to numerical solutions of these models, we have extracted values for the electron-lattice and electron-spin coupling parameters and an empirical estimate of the spin flip scattering probability for each of our samples and demonstrated the fluence-dependent variation in calculated electron, spin, and lattice temperatures as a function of the pump-probe delay. We conjecture that the increasing trend of the extracted electron-lattice coupling parameter with fluence may indicate a decrease in the efficiency of electron-lattice interactions in the low laser fluences when nonthermal electrons may play a role in influencing the magnetization dynamics. We have compared the experimental and theoretical results obtained from each of the systems under investigation and confirmed that the ratio of the Curie temperature to the atomic magnetic moment acts as a figure of merit sensitive to the demagnetization time for these systems. Our study demonstrates the mutual agreement between the phenomenological three temperature model and the microscopic three temperature model for the systems we study and highlights the possibility of a fluence-dependent enhancement in the inter-reservoir interactions in the low laser fluence regime. Finally, we have reported from our experiments a direct correlation between the ultrafast demagnetization time and the enhancement of the magnetic damping that transiently appears as a result of the laser excitation, reflecting that both phenomena are sensitive to similar underlying microscopic processes.

## 5. Acknowledgement

AB gratefully acknowledges the financial support from the S. N. Bose National Centre for Basic Sciences (SNBNCBS) under Project No. SNB/AB/18-19/211 and Department of Science and Technology (DST), Govt. of India under Grant No. DST/NM/TUE/QM-3/2019-1C-SNB. SM acknowledges DST, India for financial support from INSPIRE fellowship, while SM and SNP acknowledge SNBNCBS for senior research fellowship.

**References**

[1] C. D. Stanciu, F. Hansteen, A. V. Kimel, A. Kirilyuk, A. Tsukamoto, A. Itoh, T. Rasing, All-optical magnetic recording with circularly polarized light, Phys. Rev. Lett. 99 (2007) 047601.

[2] T. A. Ostler, J. Barker, R. F. L. Evans, R. W. Chantrell, U. Atxitia, O. Chubykalo-Fesenko, S. El Moussaoui, L. Le Guyader, E. Mengotti, L. J. Heyderman, F. Nolting, A. Tsukamoto, A. Itoh, D. Afanasiev, B. A. Ivanov, A. M. Kalashnikova, K. Vahaplar, J. Mentink, A. Kirilyuk, T. Rasing, A. V. Kimel, Ultrafast heating as a sufficient stimulus for magnetization reversal in a ferrimagnet, Nature Communications 3 (1) (2012) 666.

[3] C.-H. Lambert, S. Mangin, B. S. D. C. S. Varaprasad, Y. K. Takahashi, M. Hehn, M. Cinchetti, G. Malinowski, K. Hono, Y. Fainman, M. Aeschlimann, E. E. Fullerton, All-optical control of ferromagnetic thin films and nanostructures, Science 345 (6202) (2014) 1337–1340.

[4] C. Banerjee, N. Teichert, K. E. Siewierska, Z. Gercsi, G. Y. P. Atcheson, P. Stamenov, K. Rode, J. M. D. Coey, J. Besbas, Single pulse all-optical toggle switching of magnetization without gadolinium in the ferrimagnet mn2ruxga, Nature Communications 11 (1) (2020) 4444.

[5] T. Kampfrath, M. Battiato, P. Maldonado, G. Eilers, J. Nötzold, S. Mährlein, V. Zbarsky, F. Freimuth, Y. Mokrousov, S. Blügel, M. Wolf, I. Radu, P. M. Oppeneer, M. Münzenberg, Terahertz spin current pulses controlled by magnetic heterostructures, Nature Nanotechnology 8 (4) (2013) 256–260.

[6] Z. Jin, A. Tkach, F. Casper, V. Spetter, H. Grimm, A. Thomas, T. Kampfrath, M. Bonn, M. Kläui, D. Turchinovich, Accessing the fundamentals of magnetotransport in metals with terahertz probes, Nature Physics 11 (9) (2015) 761–766.

[7] T. Seifert, S. Jaiswal, U. Martens, J. Hannegan, L. Braun, P. Maldonado, F. Freimuth, A. Kronenberg, J. Henrizi, I. Radu, E. Beaurepaire, Y. Mokrousov, P. M. Oppeneer, M. Jourdan, G. Jakob, D. Turchinovich, L. M. Hayden, M. Wolf, M. Münzenberg, M. Kläui, T. Kampfrath, Efficient metallic spintronic emitters of ultrabroadband terahertz radiation, Nature

Photonics 10 (7) (2016) 483–488.

[8] A. Vaterlaus, T. Beutler, F. Meier, Spin-lattice relaxation time of ferromagnetic gadolinium determined with time-resolved spin-polarized photoemission, Phys. Rev. Lett. 67 (1991) 3314–3317.

[9] J. Hohlfeld, E. Matthias, R. Knorren, K. H. Bennemann, Nonequilibrium magnetization dynamics of nickel [phys. rev. lett. 78, 4861 (1997)], Phys. Rev. Lett. 79 (1997) 960–960.

[10] E. Beaurepaire, M. Maret, V. Halté, J.-C. Merle, A. Daunois, J.-Y. Bigot, Spin dynamics in $copt_3$ alloy films: A magnetic phase transition in the femtosecond time scale, Phys. Rev. B 58 (1998) 12134–12137.

[11] E. Beaurepaire, J.-C. Merle, A. Daunois, J.-Y. Bigot, Ultrafast spin dynamics in ferromagnetic nickel, Phys. Rev. Lett. 76 (1996) 4250–4253.

[12] M. Aeschlimann, M. Bauer, S. Pawlik, W. Weber, R. Burgermeister, D. Oberli, H. C. Siegmann, Ultrafast spin-dependent electron dynamics in fcc co, Phys. Rev. Lett. 79 (1997) 5158–5161.

[13] B. Koopmans, M. van Kampen, J. T. Kohlhepp, W. J. M. de Jonge, Ultrafast magneto-optics in nickel: Magnetism or optics?, Phys. Rev. Lett. 85 (2000) 844–847.

[14] T. Kampfrath, R. G. Ulbrich, F. Leuenberger, M. Münzenberg, B. Sass, W. Felsch, Ultrafast magneto-optical response of iron thin films, Phys. Rev. B 65 (2002) 104429.

[15] H.-S. Rhie, H. A. Dürr, W. Eberhardt, Femtosecond electron and spin dynamics in Ni/W(110) films, Phys. Rev. Lett. 90 (2003) 247201.

[16] B. Koopmans, J. J. M. Ruigrok, F. D. Longa, W. J. M. de Jonge, Unifying ultrafast magnetization dynamics, Phys. Rev. Lett. 95 (2005) 267207.

[17] M. Cinchetti, M. Sánchez Albaneda, D. Hoffmann, T. Roth, J.-P. Wüstenberg, M. Krauß, O. Andreyev, H. C. Schneider, M. Bauer, M. Aeschlimann, Spin-flip processes and ultrafast magnetization dynamics in co: Unifying the microscopic and macroscopic view of femtosecond magnetism, Phys. Rev. Lett. 97 (2006) 177201.

[18] J. Walowski, G. Müller, M. Djordjevic, M. Münzenberg, M. Kläui, C. A. F. Vaz, J. A. C. Bland, Energy equilibration processes of electrons, magnons, and phonons at the femtosecond time scale, Phys. Rev. Lett. 101 (2008) 237401.

[19] I. Radu, G. Woltersdorf, M. Kiessling, A. Melnikov, U. Bovensiepen, J.-U. Thiele, C. H. Back, Laser-induced magnetization dynamics of lanthanide-doped permalloy thin films, Phys. Rev. Lett. 102 (2009) 117201.

[20] M. Battiato, K. Carva, P. M. Oppeneer, Superdiffusive spin transport as a mechanism of ultrafast demagnetization, Phys. Rev. Lett. 105 (2010) 027203.

[21] B. Vodungbo, J. Gautier, G. Lambert, A. B. Sardinha, M. Lozano, S. Sebban, M. Ducousso, W. Boutu, K. Li, B. Tudu, M. Tortarolo, R. Hawaldar, R. Delaunay, V. López-Flores, J. Arabski, C. Boeglin, H. Merdji, P. Zeitoun, J. Lüning, Laser-induced ultrafast demagnetization in the presence of a nanoscale magnetic domain network, Nature Communications 3 (1) (2012) 999.

[22] S. Pan, O. Hellwig, A. Barman, Controlled coexcitation of direct and indirect ultrafast demagnetization in co/pd multilayers with large perpendicular magnetic anisotropy, Phys. Rev. B 98 (2018) 214436.

[23] B. Koopmans, G. Malinowski, F. Dalla Longa, D. Steiauf, M. Fähnle, T. Roth, M. Cinchetti, M. Aeschlimann, Explaining the paradoxical diversity of ultrafast laser-induced demagnetization, Nature Materials 9 (3) (2010) 259–265.

[24] A. J. Schellekens, W. Verhoeven, T. N. Vader, B. Koopmans, Investigating the contribution of superdiffusive transport to ultrafast demagnetization of ferromagnetic thin films, Applied Physics

Letters 102 (25) (2013) 252408.

[25] G. Zhang, W. Hübner, E. Beaurepaire, J.-Y. Bigot, Laser-Induced Ultrafast Demagnetization: Femtomagnetism, a New Frontier?, Springer Berlin Heidelberg, Berlin, Heidelberg, 2002, pp. 245–289.

[26] C. Stamm, T. Kachel, N. Pontius, R. Mitzner, T. Quast, K. Holldack, S. Khan, C. Lupulescu, E. F. Aziz, M. Wietstruk, H. A. Dürr, W. Eberhardt, Femtosecond modification of electron localization and transfer of angular momentum in nickel, Nature Materials 6 (10) (2007) 740–743.

[27] J. Walowski, M. Münzenberg, Perspective: Ultrafast magnetism and thz spintronics, Journal of Applied Physics 120 (14) (2016) 140901.

[28] G. Malinowski, F. Dalla Longa, J. H. H. Rietjens, P. V. Paluskar, R. Huijink, H. J. M. Swagten, B. Koopmans, Control of speed and efficiency of ultrafast demagnetization by direct transfer of spin angular momentum, Nature Physics 4 (11) (2008) 855–858.

[29] W. Hübner, G. P. Zhang, Ultrafast spin dynamics in nickel, Phys. Rev. B 58 (1998) R5920–R5923.

[30] A. Barman, J. Sinha, Spin dynamics and damping in ferromagnetic thin films and nanostructures, Springer International Publishing, 2018.

[31] S. Pan, S. Mondal, T. Seki, K. Takanashi, A. Barman, Influence of thickness-dependent structural evolution on ultrafast magnetization dynamics in $Co_2Fe_{0.4}Mn_{0.6}Si$ heusler alloy thin films, Phys. Rev. B 94 (2016) 184417.

[32] C. La-O-Vorakiat, E. Turgut, C. A. Teale, H. C. Kapteyn, M. M. Murnane, S. Mathias, M. Aeschlimann, C. M. Schneider, J. M. Shaw, H. T. Nembach, T. J. Silva, Ultrafast demagnetization measurements using extreme ultraviolet light: Comparison of electronic and magnetic contributions, Phys. Rev. X 2 (2012) 011005.

[33] B. Mansart, D. Boschetto, A. Savoia, F. Rullier-Albenque, F. Bouquet, E. Papalazarou, A. Forget, D. Colson, A. Rousse, M. Marsi, Ultrafast transient response and electron- phonon coupling in the iron-pnictide superconductor $Ba(Fe_{1-x}Co_x)_2as_2$, Phys. Rev. B 82 (2010) 024513.

[34] V. Shokeen, M. Sanchez Piaia, J.-Y. Bigot, T. Müller, P. Elliott, J. K. Dewhurst, S. Sharma, E. K. U. Gross, Spin flips versus spin transport in nonthermal electrons excited by ultrashort optical pulses in transition metals, Phys. Rev. Lett. 119 (2017) 107203.

[35] L. G. Schulz, The optical constants of silver, gold, copper, and aluminum. i. the absorption coefficient k, J. Opt. Soc. Am. 44 (5) (1954) 357–362.

[36] E. Carpene, E. Mancini, C. Dallera, M. Brenna, E. Puppin, S. De Silvestri, Dynamics of electron-magnon interaction and ultrafast demagnetization in thin iron films, Phys. Rev. B 78 (2008) 174422.

[37] F. Dalla Longa, J. T. Kohlhepp, W. J. M. de Jonge, B. Koopmans, Influence of photon angular momentum on ultrafast demagnetization in nickel, Phys. Rev. B 75 (2007) 224431.

[38] K. C. Kuiper, G. Malinowski, F. D. Longa, B. Koopmans, Nonlocal ultrafast magnetization dynamics in the high fluence limit, Journal of Applied Physics 109 (7) (2011) 07D316.

[39] T. Roth, A. J. Schellekens, S. Alebrand, O. Schmitt, D. Steil, B. Koopmans, M. Cinchetti, M. Aeschlimann, Temperature dependence of laser-induced demagnetization in ni: A key for identifying the underlying mechanism, Phys. Rev. X 2 (2012) 021006.

[40] M. Krauß, T. Roth, S. Alebrand, D. Steil, M. Cinchetti, M. Aeschlimann, H. C. Schneider, Ultrafast demagnetization of ferromagnetic transition metals: The role of the coulomb interaction, Phys. Rev. B 80 (2009) 180407.

[41] S. Eich, M. Plötzing, M. Rollinger, S. Emmerich, R. Adam, C. Chen, H. C. Kapteyn, M. M. Murnane, L. Plucinski, D. Steil, B. Stadtmüller, M. Cinchetti, M. Aeschlimann, C. M. Schneider, S. Mathias, Band structure evolution during the ultrafast ferromagnetic- paramagnetic phase transition in cobalt, Science Advances 3 (3) (2017) e1602094.

[42] C. Kittel, Introduction to solid state physics, Progress in Materials Science (2004).

[43] M. Braun, R. Kohlhaas, The specific heat of cobalt between 50 and 1400°c, Zeit f. Naturforsch. 19a (1964) 663.

[44] N. Ashcroft, N. Mermin, Solid State Physics, Cengage Learning, 2011.

[45] P.-W. Ma, S. L. Dudarev, C. H. Woo, Spin-lattice-electron dynamics simulations of magnetic materials, Phys. Rev. B 85 (2012) 184301.

[46] P. J. Meschter, J. W. Wright, C. R. Brooks, T. G. Kollie, Physical contributions to the heat capacity of nickel, Journal of Physics and Chemistry of Solids 42 (9) (1981) 861–871.

[47] W. Betteridge, The properties of metallic cobalt, Progress in Materials Science 24 (1980) 51–142.

[48] A. J. Schellekens, B. Koopmans, Comparing ultrafast demagnetization rates between competing models for finite temperature magnetism, Phys. Rev. Lett. 110 (2013) 217204.

[49] A. Eschenlohr, M. Battiato, P. Maldonado, N. Pontius, T. Kachel, K. Holldack, R. Mitzner, A. Föhlisch, P. M. Oppeneer, C. Stamm, Ultrafast spin transport as key to femtosecond demagnetization, Nature Materials 12 (4) (2013) 332–336.

[50] B. Y. Mueller, B. Rethfeld, Relaxation dynamics in laser-excited metals under nonequilibrium conditions, Phys. Rev. B 87 (2013) 035139.

[51] E. G. Tveten, A. Brataas, Y. Tserkovnyak, Electron-magnon scattering in magnetic heterostructures far out of equilibrium, Phys. Rev. B 92 (2015) 180412.

[52] A. B. Schmidt, M. Pickel, M. Donath, P. Buczek, A. Ernst, V. P. Zhukov, P. M. Echenique, L. M. Sandratskii, E. V. Chulkov, M. Weinelt, Ultrafast magnon generation in an fe film on cu(100), Phys. Rev. Lett. 105 (2010) 197401.

[53] S. Mathias, C. La-O-Vorakiat, P. Grychtol, P. Granitzka, E. Turgut, J. M. Shaw, R. Adam, H. T. Nembach, M. E. Siemens, S. Eich, C. M. Schneider, T. J. Silva, M. Aeschlimann, M. M. Murnane, H. C. Kapteyn, Probing the timescale of the exchange interaction in a ferromagnetic alloy, Proceedings of the National Academy of Sciences 109 (13) (2012) 4792–4797.

[54] I. Radu, C. Stamm, A. Eschenlohr, F. Radu, R. Abrudan, K. Vahaplar, T. Kachel, N. Pontius, R. Mitzner, K. Holldack, A. Föhlisch, T. A. Ostler, J. H. Mentink, R. F. L. Evans, R. W. Chantrell, A. Tsukamoto, A. Itoh, A. Kirilyuk, A. V. Kimel, T. Rasing, Ultrafast and distinct spin dynamics in magnetic alloys, SPIN 05 (03) (2015) 1550004.

[55] W. Zhang, W. He, X.-Q. Zhang, Z.-H. Cheng, J. Teng, M. Fähnle, Unifying ultrafast demagnetization and intrinsic gilbert damping in co/ni bilayers with electronic relaxation near the fermi surface, Phys. Rev. B 96 (2017) 220415.

[56] W. Zhang, Q. Liu, Z. Yuan, K. Xia, W. He, Q.-f. Zhan, X.-q. Zhang, Z.-h. Cheng, Enhancement of ultrafast demagnetization rate and gilbert damping driven by femtosecond laser-induced spin currents in $Fe_{81}Ga_{19}/Ir_{20}Mn_{80}$ bilayers, Phys. Rev. B 100 (2019) 104412.

[57] B. Liu, X. Ruan, Z. Wu, H. Tu, J. Du, J. Wu, X. Lu, L. He, R. Zhang, Y. Xu, Transient enhancement of magnetization damping in cofeb film via pulsed laser excitation, Applied Physics Letters 109 (4) (2016) 042401.

[58] V. Kamberský, On the landau–lifshitz relaxation in ferromagnetic metals, Canadian Journal of Physics 48 (24) (1970) 2906–2911.

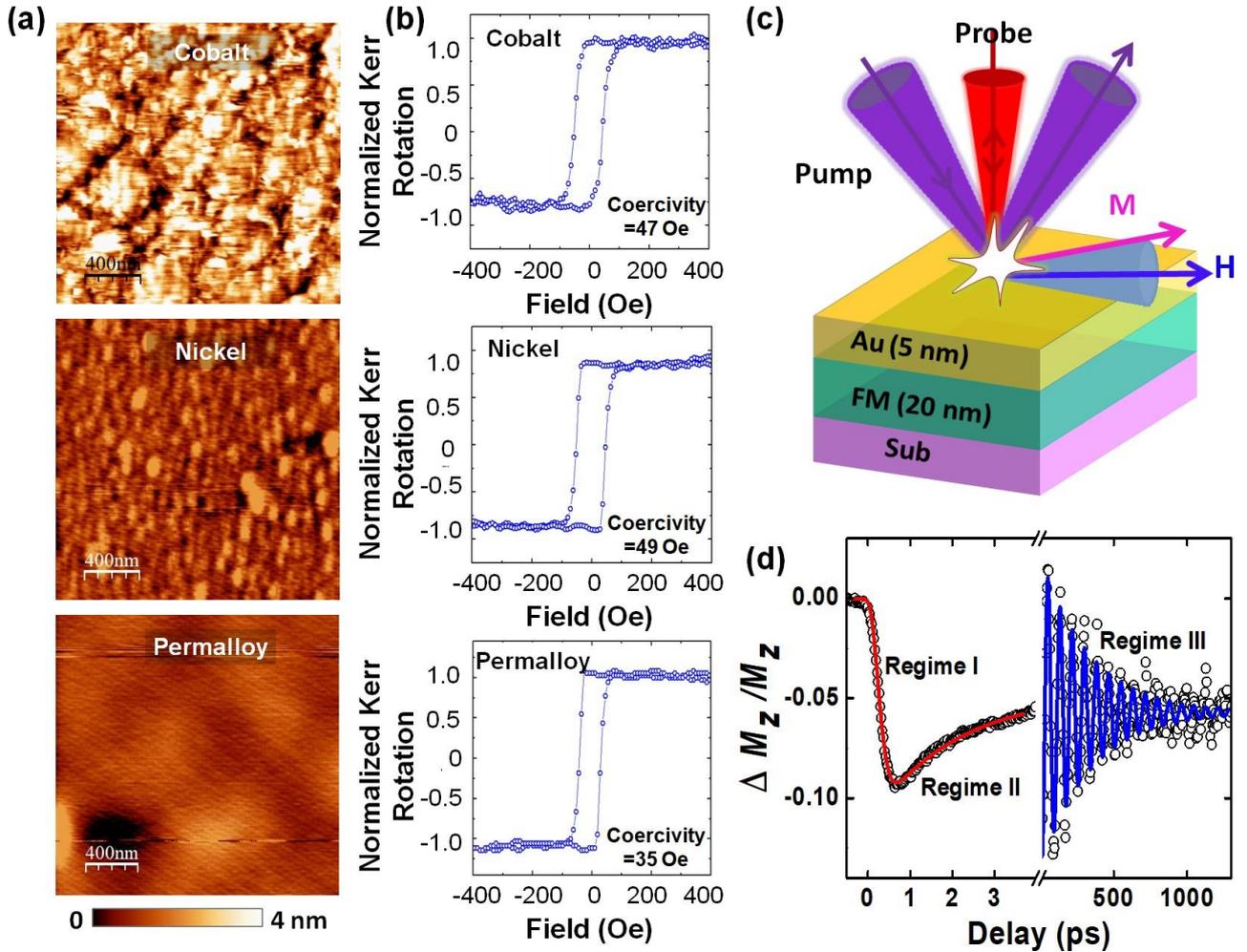

Figure 1: (*a*) Atomic force microscopy images of the samples showing the surface topography. (*b*) Static-MOKE hysteresis loops measured for each sample. (*c*) A schematic of the samples used in our measurements and the measurement geometry. The 400 nm pump and 800 nm probe are incident non-collinearly on the sample surface.
(*d*) Typical TR-MOKE data showing different temporal regimes of the magnetization dynamics.

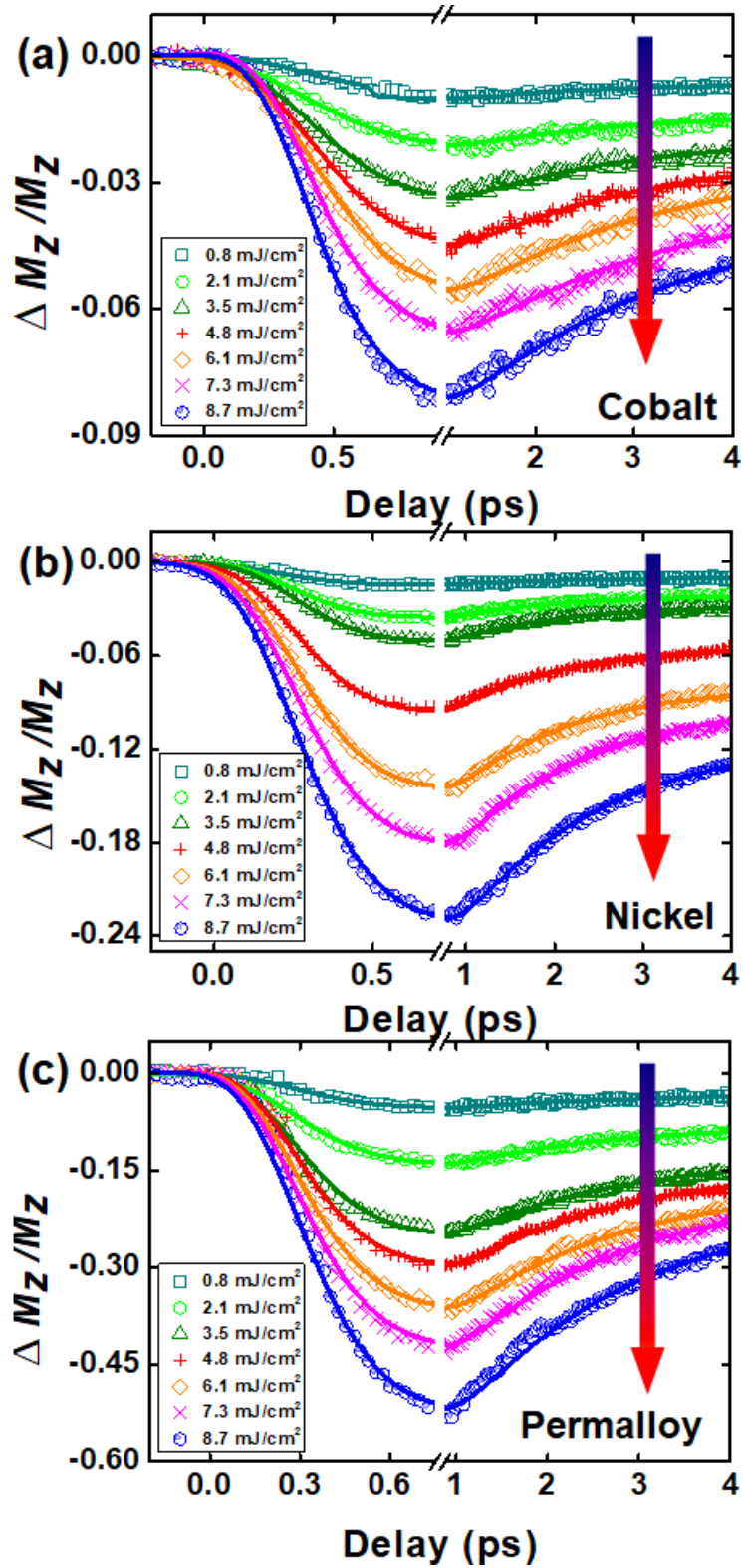

Figure 2: Ultrafast demagnetization traces obtained from TR-MOKE measurements on (*a*) cobalt, (*b*) nickel and (*c*) permalloy thin films at various pump fluences. The downward arrows represent increased pump fluence. The symbols represent experimental data points and the solid lines represent fit to Equation 3.

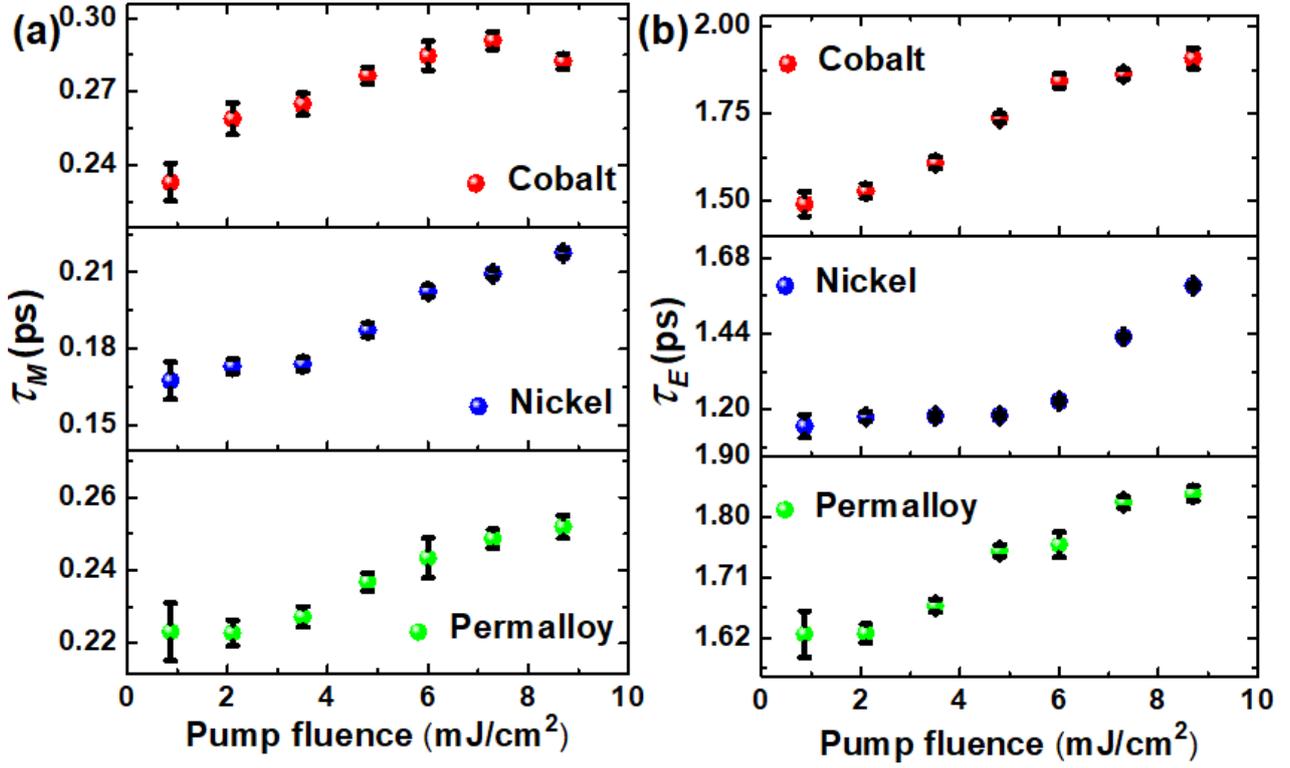

Figure 3: Variation of (*a*) demagnetization time $\tau_M$ and (*b*) fast relaxation time $\tau_E$ with the pump fluence for 20-nm-thick cobalt, nickel and permalloy films. Both timescales show an increasing trend with increasing pump fluence.

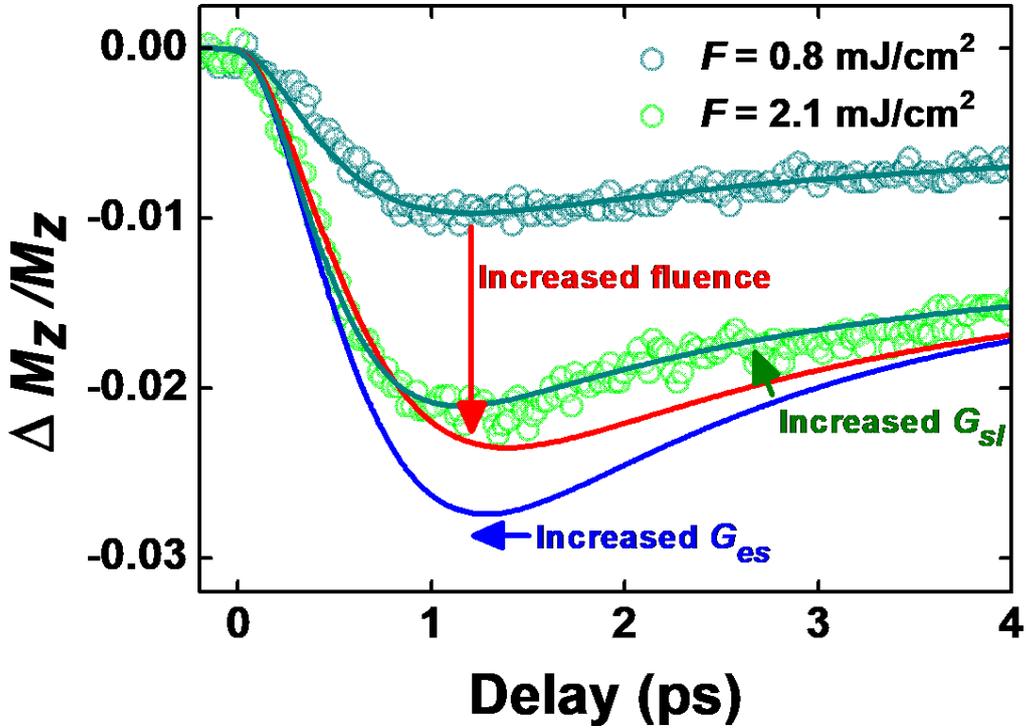

Figure 4: Sensitivity of the demagnetization traces to the various adjustable fit parameters. Symbols represent data points and solid lines represent fits to the 3TM. Data shown is for the cobalt thin film. For $F =0.8$ mJ/cm$^2$, $G_{el}=8.2 \times 10^{17}$ Js$^{-1}$m$^{-3}$K$^{-1}$ and $G_{es}=6.9 \times 10^{16}$ Js$^{-1}$m$^{-3}$K$^{-1}$; for $F =2.1$ mJ/cm$^2$, $G_{el}=1.1 \times 10^{18}$ Js$^{-1}$m$^{-3}$K$^{-1}$ and $G_{es}=8 \times 10^{17}$ Js$^{-1}$m$^{-3}$K$^{-1}$.

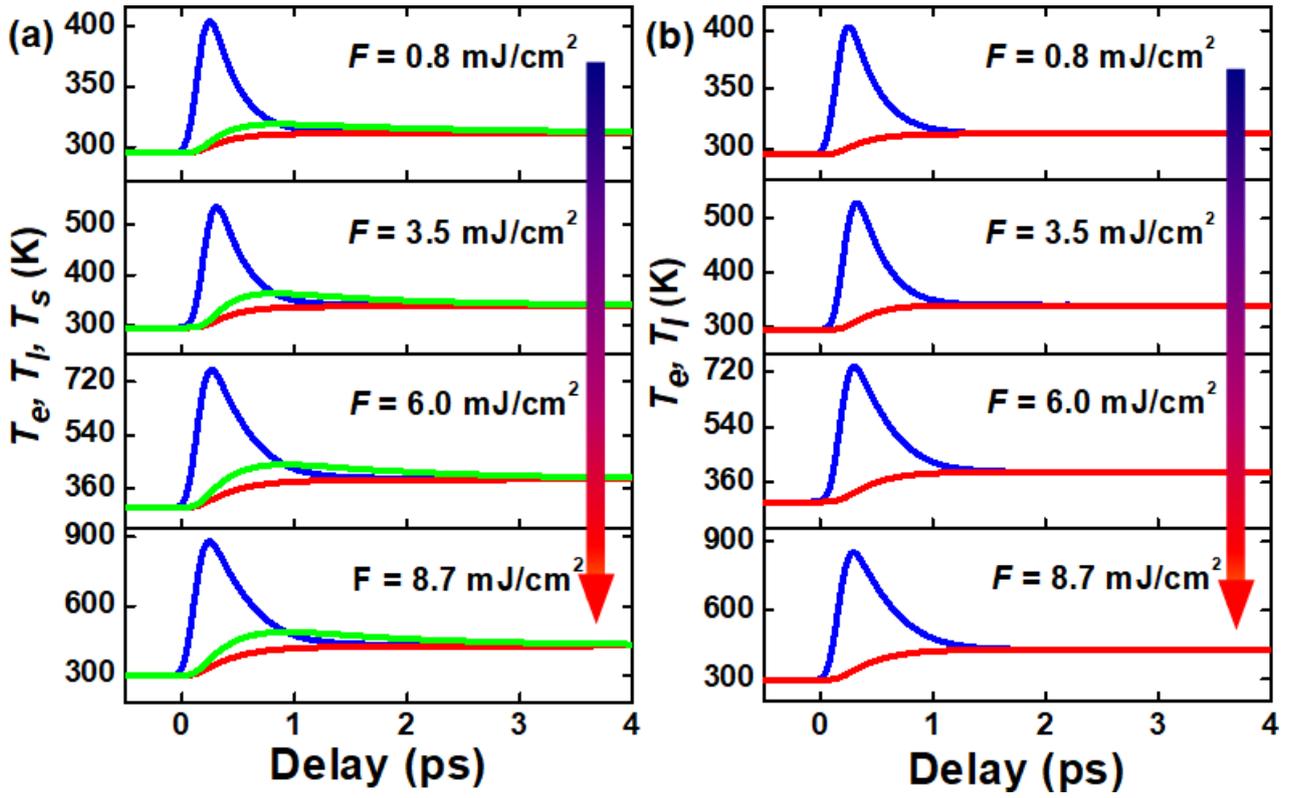

Figure 5: Temporal evolution of the reservoir temperatures of nickel $T_e$ (blue), $T_l$ (red), and $T_s$ (green) calculated using the (a) phenomenological three temperature model and (b) the microscopic three temperature model.

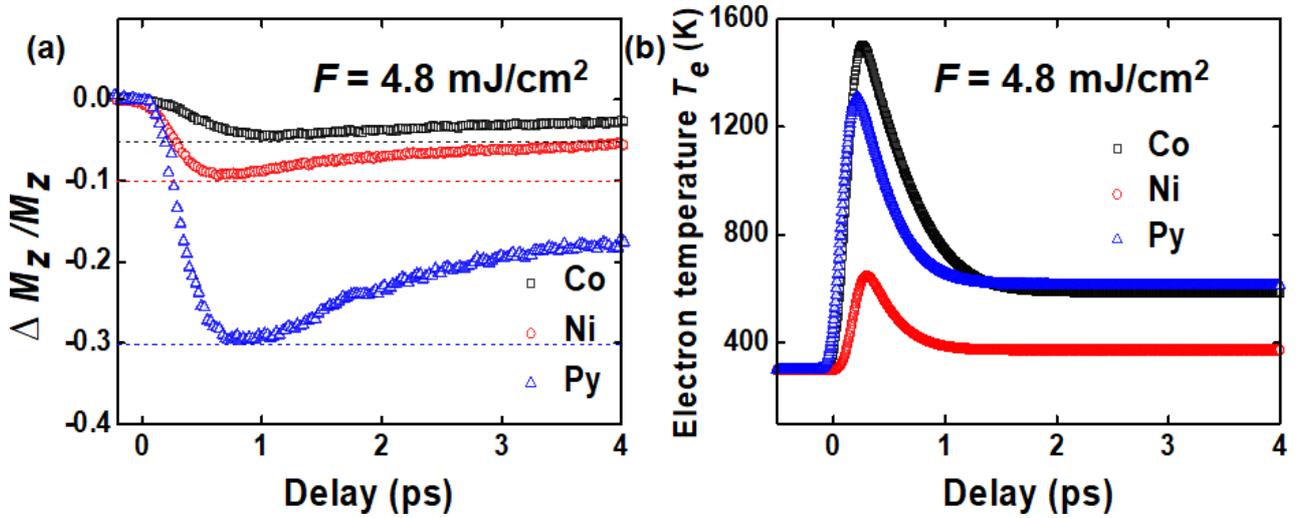

Figure 6: Comparison between (a) the experimental demagnetization and (b) the simulated electron temperature profiles for cobalt, nickel and permalloy at a pump fluence of 4.8 mJ/cm$^2$.

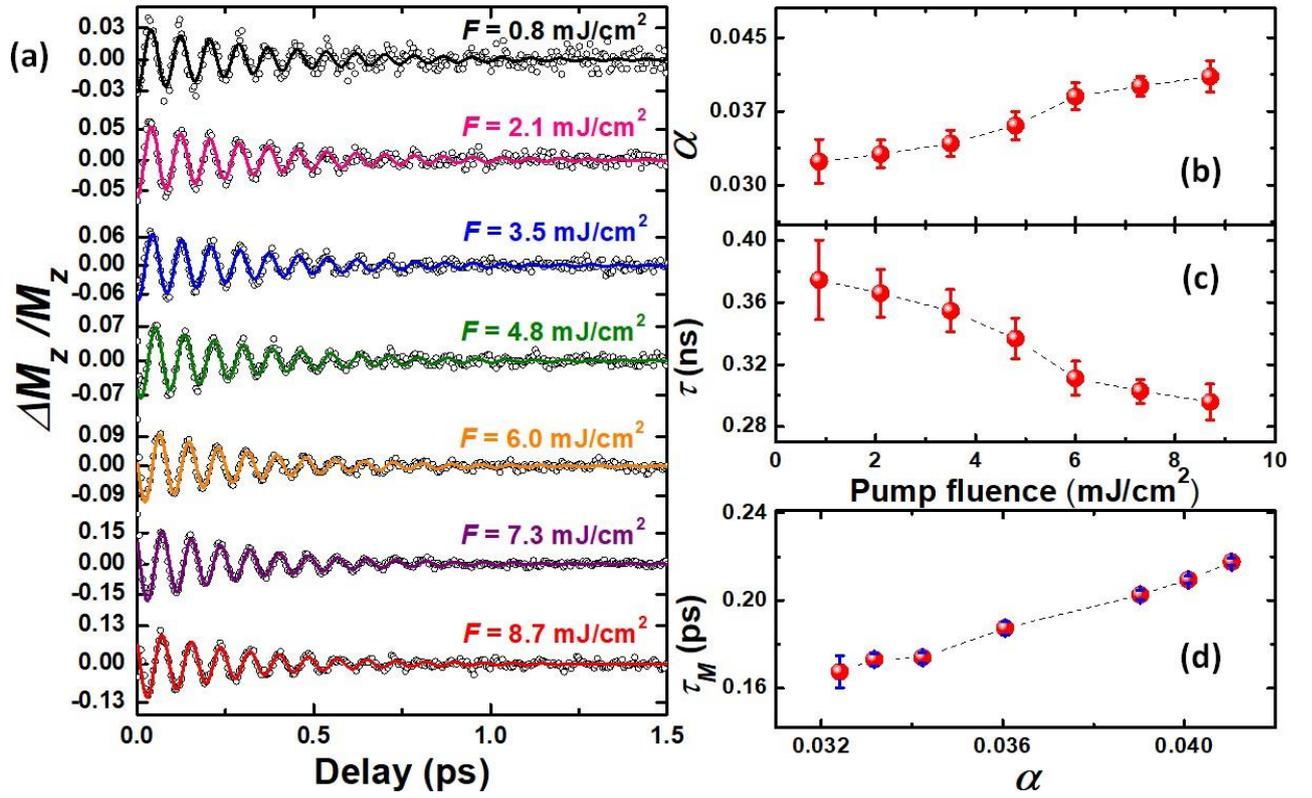

Figure 7: (*a*) Fluence-dependent variation of precessional dynamics for the nickel thin film. Symbols are experimental data points and the solid lines are fits to Equation 5. (*b*) Variation of $f$ with pump fluence. (*c*) Variation of $\alpha$ with pump fluence. (*d*) Direct correlation between $\tau_M$ and $\alpha$.

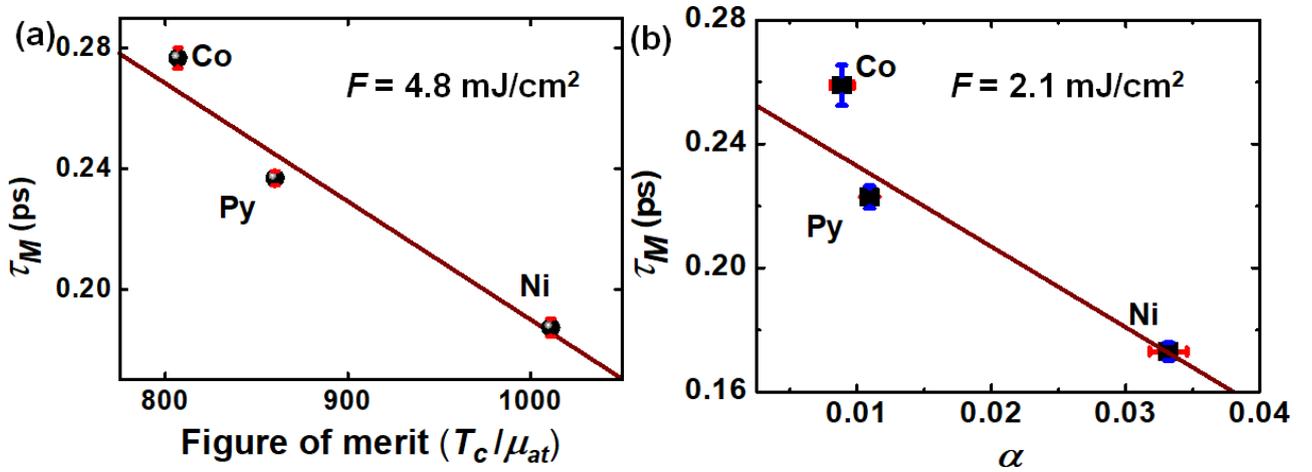

Figure 8: (*a*) Inverse variation of the demagnetization time with the ratio of Curie temperature to atomic magnetic moment and (*b*) inverse correlation of the demagnetization time with magnetic damping across three different materials.

**Supplementary Materials**

# I. Transient reflectivity measurements

Time-resolved transient reflectivity signal is recorded by TR-MOKE measurements simultaneously with the Kerr rotation signal quantifying the magnetic response. At any given fluence, the maximum amplitude of the transient reflectivity change is about a factor of 100 smaller as compared to that of the Kerr rotation. This validates that the Kerr rotation signal reflects the genuine magnetic response with negligible nonmagnetic contribution.

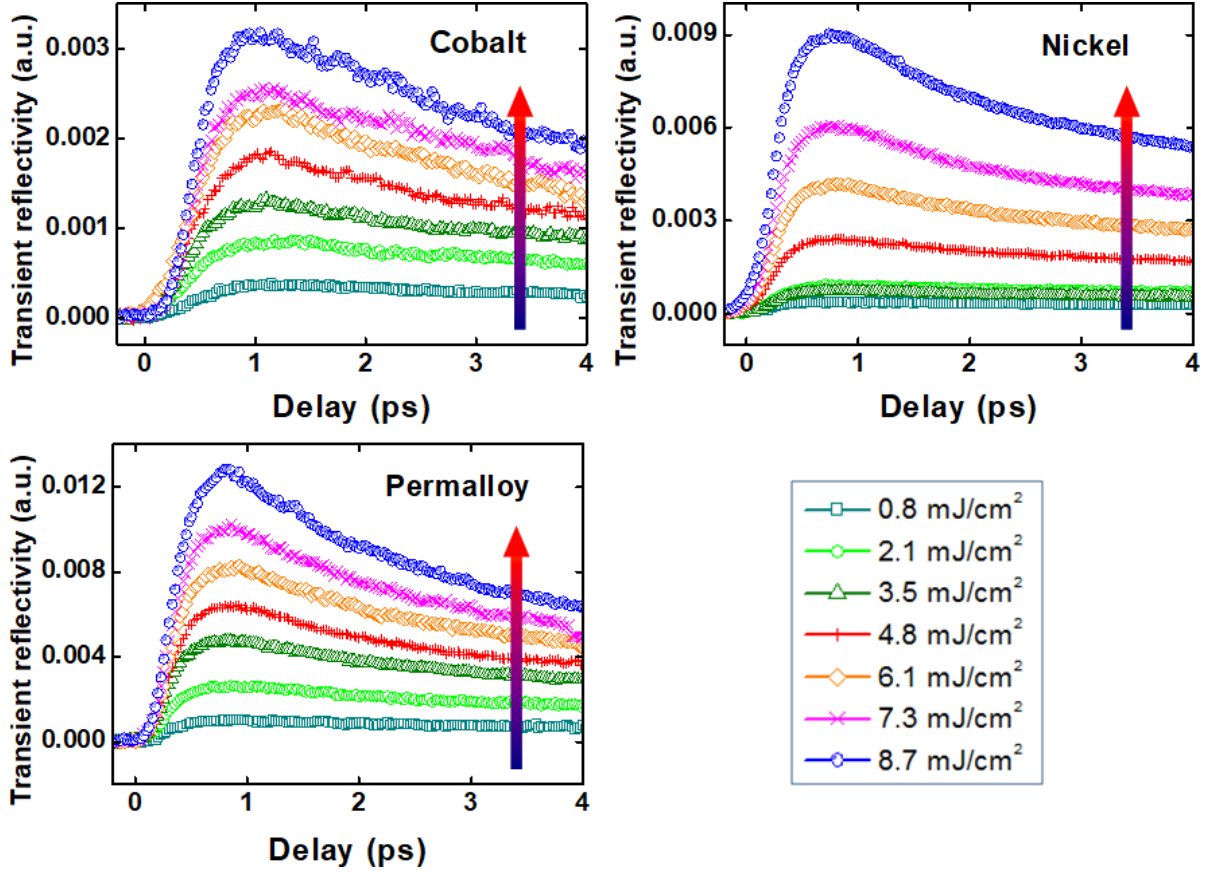

Figure S1: Time-resolved transient reflectivity measured by TR-MOKE for cobalt, nickel, and permalloy. The upward arrows represent increased pump fluence.

# II. Calculation of electron, spin, and lattice temperature profiles

The calculated temporal evolution of the reservoir temperatures for the 20-nm-thick cobalt and permalloy thin films are represented in Fig. S2 and Fig. S3, respectively. The electron temperature shows a rapid and dramatic increase within one picosecond of the laser excitation, beyond which it decays to an equilibrium value over a few picoseconds. Over this period, the lattice subsystem is gradually heated above the equilibrium temperature while the heating of the spin subsystem directly leads to the demagnetization. The maximum electron, spin and lattice temperatures increase with increasing laser fluence.

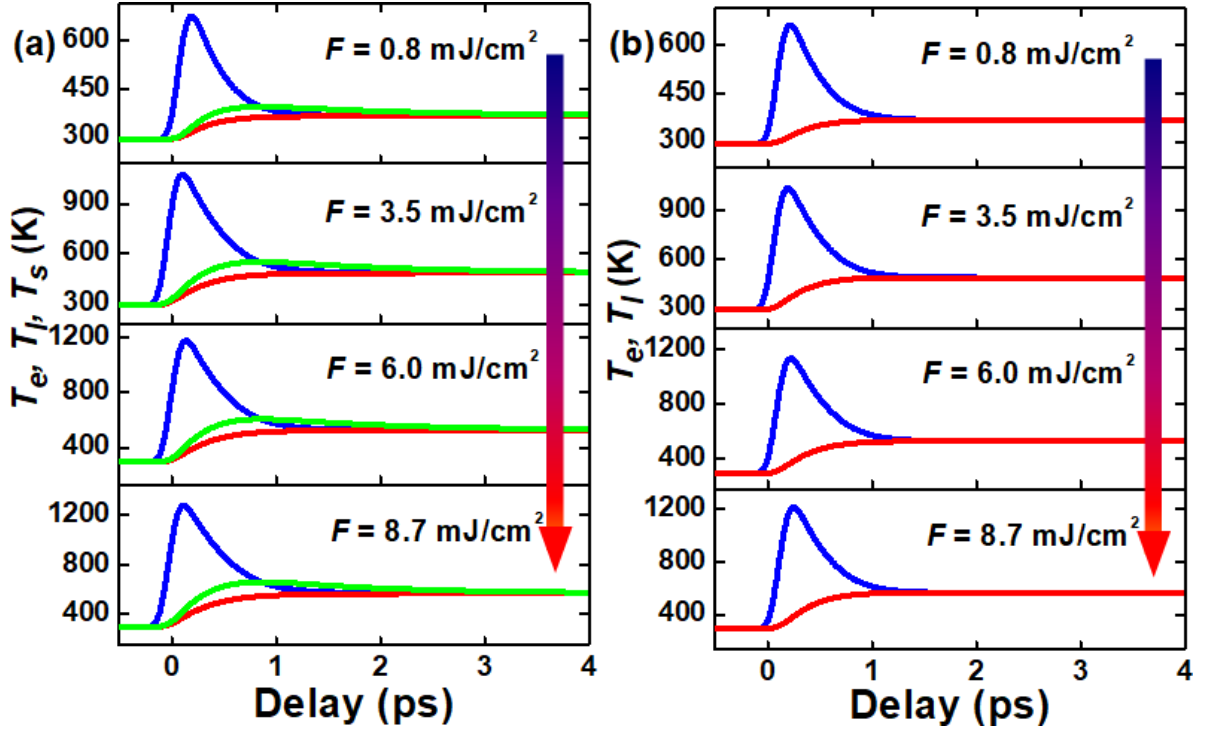

Figure S2: Electron, lattice, and spin temperature profiles for the cobalt thin film calculated using (a) the three temperature model and (b) the microscopic three temperature model for various laser fluences (blue: $T_e$, dashed, red: $T_l$ and dotted, green: $T_s$).

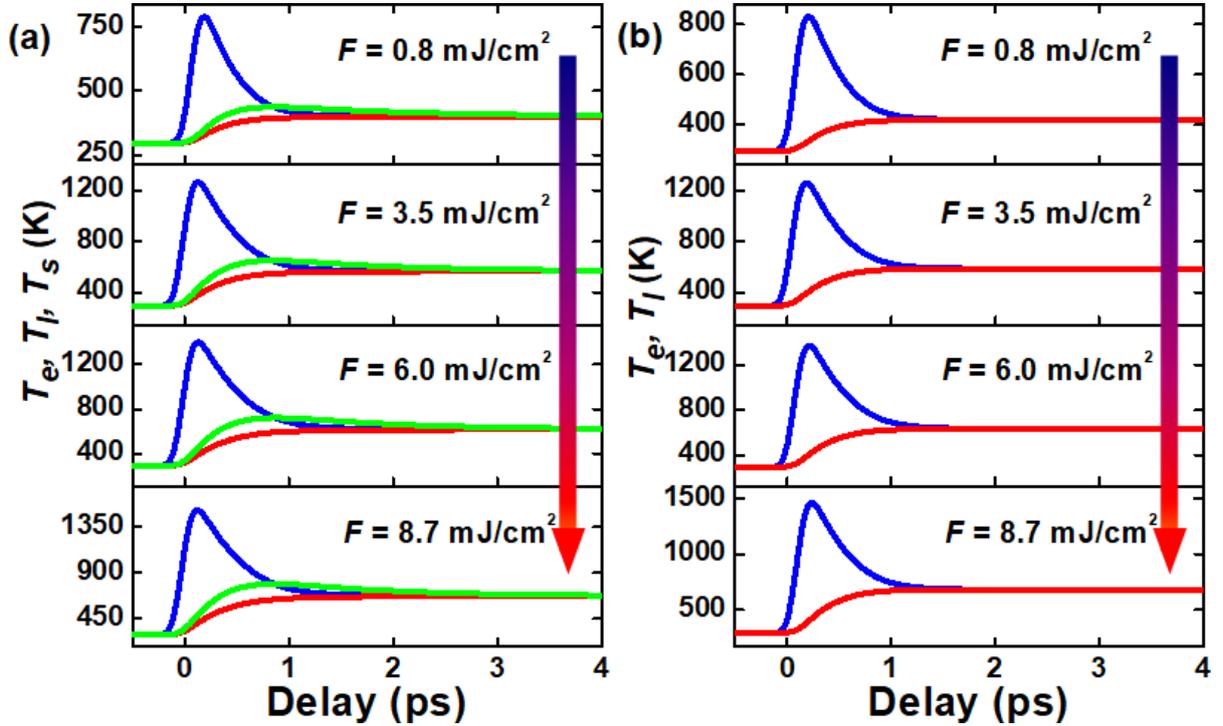

Figure S3: Electron, lattice, and spin temperature profiles for the permalloy thin film calculated using (a) the three temperature model and (b) the microscopic three temperature model for various laser fluences).

### III.  Extraction of inter-reservoir coupling constants

The least-squares fitting routines set up based on the three temperature model and the microscopic three temperature model return the values of the inter-reservoir coupling constants that best reproduce the experimental demagnetization traces. The microscopic three temperature model additionally returns the value of spin-flip probability $a_{sf}$ associated with the spin-flip scattering events which directly lead to the magnetization quenching. The extracted values of these microscopic parameters are given in Tables S1-S3. Fig. S4 shows a comparison of the values of the electron-lattice coupling parameter $G_{el}$ as extracted from the fits to the three temperature model and the microscopic three temperature model. For reasons discussed in the main text, the spin-lattice coupling parameter $G_{sl}$ is taken to be zero for the three temperature model simulations. Fig. S5 shows that, when $G_{sl}$ is set equal to $3 \times 10^{16}$ Jm$^{-1}$s$^{-1}$m$^{-3}$K$^{-1}$ as considered in *Phys. Rev. Lett.* 76, 4250 (1996), the simulated trace and calculated electron temperature profile are nearly indistinguishable from the case in which $G_{sl}$ is neglected.

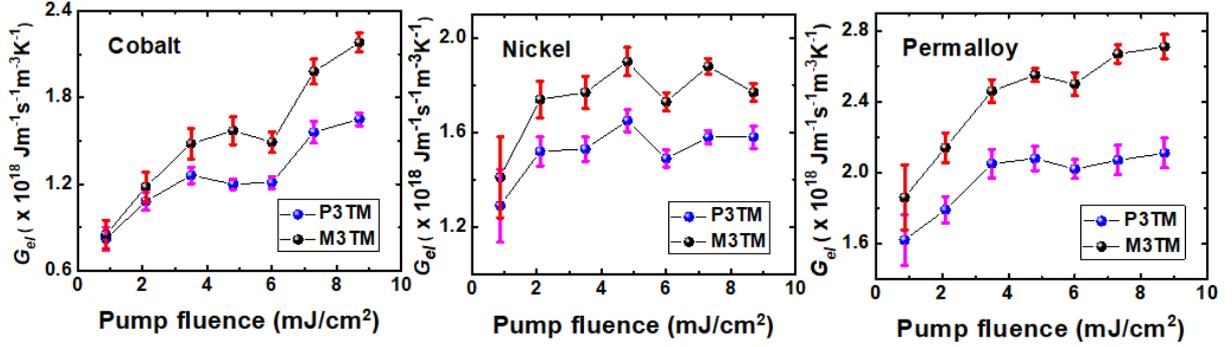

Figure S4: Values of the electron-lattice coupling parameter $G_{el}$ extracted from fitting to the three temperature model and microscopic three temperature model.

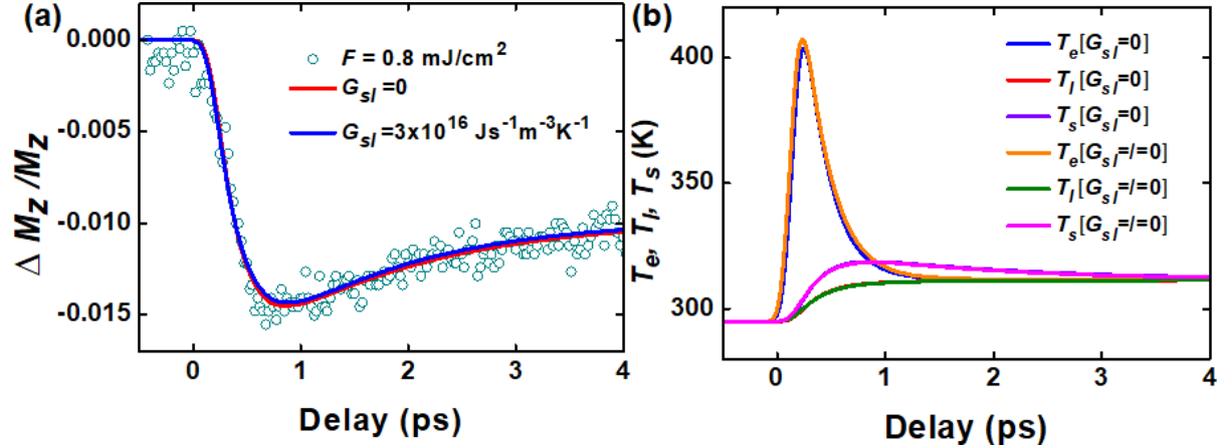

Figure S5: (a) Results of three temperature modelling and (b) calculated reservoir temperature profiles for two values of $G_{sl}$.

| Pump fluence (mJ/cm$^2$) | $G_{el[3TM]}$ ($\times 10^{18}$) (Js$^{-1}$m$^{-3}$K$^{-1}$) | $G_{el[M3TM]}$ ($\times 10^{18}$) (Js$^{-1}$m$^{-3}$K$^{-1}$) | $G_{es}$ ($\times 10^{17}$) (Js$^{-1}$m$^{-3}$K$^{-1}$) | $a_{sf}$ ($\times 10^{-1}$) |
|---|---|---|---|---|
| 0.8 | 0.82±0.081 | 0.85±0.097 | 0.69±0.037 | 0.19±0.023 |
| 2.1 | 1.08±0.063 | 1.18±0.105 | 0.8±0.029 | 0.16±0.014 |
| 3.5 | 1.26±0.057 | 1.48±0.106 | 0.91±0.028 | 0.16±0.009 |
| 4.8 | 1.2±0.037 | 1.57±0.096 | 0.97±0.015 | 0.17±0.008 |
| 6.0 | 1.21±0.042 | 1.49±0.072 | 1.02±0.022 | 0.17±0.008 |
| 7.3 | 1.56±0.076 | 1.98±0.087 | 1.15±0.041 | 0.16±0.004 |
| 8.7 | 1.65±0.045 | 2.18±0.066 | 1.28±0.036 | 0.17±0.003 |

Table S1: Microscopic parameters extracted from the 3TM and M3TM simulations for cobalt.

| Pump fluence (mJ/cm$^2$) | $G_{el[3TM]}$ ($\times 10^{18}$) (Js$^{-1}$m$^{-3}$K$^{-1}$) | $G_{el[M3TM]}$ ($\times 10^{18}$) (Js$^{-1}$m$^{-3}$K$^{-1}$) | $G_{es}$ ($\times 10^{17}$) (Js$^{-1}$m$^{-3}$K$^{-1}$) | $a_{sf}$ ($\times 10^{-1}$) |
|---|---|---|---|---|
| 0.8 | 1.29±0.154 | 1.41±0.172 | 2.1±0.182 | 0.53±0.03 |
| 2.1 | 1.52±0.061 | 1.74±0.077 | 2.84±0.087 | 0.55±0.012 |
| 3.5 | 1.53±0.051 | 1.77±0.069 | 2.93±0.076 | 0.56±0.011 |
| 4.8 | 1.65±0.047 | 1.9±0.06 | 2.86±0.064 | 0.51±0.008 |
| 6.0 | 1.49±0.036 | 1.73±0.039 | 2.74±0.048 | 0.57±0.007 |
| 7.3 | 1.58±0.028 | 1.88±0.033 | 2.87±0.038 | 0.59±0.006 |
| 8.7 | 1.58±0.048 | 1.77±0.039 | 2.77±0.062 | 0.62±0.008 |

Table S2: Microscopic parameters extracted from the 3TM and M3TM simulations for nickel.

| Pump fluence (mJ/cm$^2$) | $G_{el[3TM]}$ ($\times 10^{18}$) (Js$^{-1}$m$^{-3}$K$^{-1}$) | $G_{el[M3TM]}$ ($\times 10^{18}$) (Js$^{-1}$m$^{-3}$K$^{-1}$) | $G_{es}$ ($\times 10^{17}$) (Js$^{-1}$m$^{-3}$K$^{-1}$) | $a_{sf}$ ($\times 10^{-1}$) |
|---|---|---|---|---|
| 0.8 | 1.62±0.144 | 1.86±0.184 | 2.16±0.145 | 0.31±0.016 |
| 2.1 | 1.79±0.074 | 2.14±0.084 | 2.34±0.073 | 0.34±0.007 |
| 3.5 | 2.05±0.081 | 2.46±0.065 | 2.65±0.084 | 0.39±0.005 |
| 4.8 | 2.08±0.068 | 2.55±0.038 | 2.81±0.071 | 0.45±0.006 |
| 6.0 | 2.02±0.053 | 2.5±0.065 | 2.72±0.058 | 0.49±0.007 |
| 7.3 | 2.07±0.083 | 2.67±0.052 | 2.87±0.089 | 0.56±0.009 |
| 8.7 | 2.11±0.084 | 2.71±0.069 | 2.87±0.086 | 0.64±0.01 |

Table S1: Microscopic parameters extracted from the 3TM and M3TM simulations for permalloy.

### IV. Temperature dependence of spin specific heat

Considering the theory of Manchon et al. (*Phys. Rev. B* 85, 064408 (2012)) as adopted in *Sci. Rep.* 10:6355 (2020), the specific heat $C_s$ of the spin subsystem in the 3TM framework can be calculated as a function of the spin temperature $T_s$ as:

$$C_s(T_s) = \left(C_0 \times \ln \frac{1+(T_s/T_c)^3}{1-(T_s/T_c)^3}\right) + C_0 \qquad (S1)$$

Fig. S6 below shows the trends in the inter-reservoir coupling parameters for two scenarios: considering the room temperature value of $C_s(T_0)$ or an explicit temperature (and hence time) dependence of $C_s$.

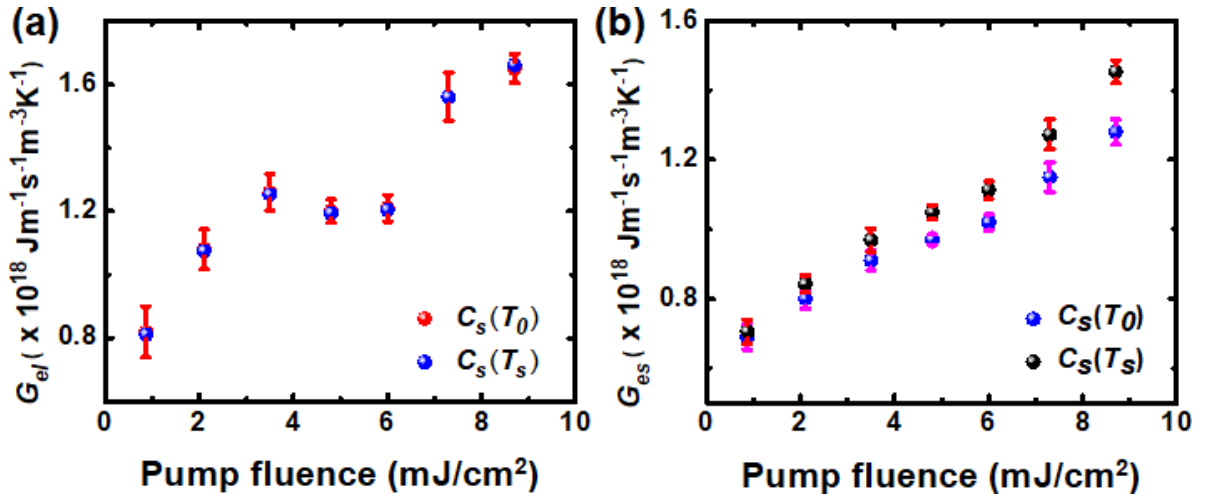

Figure S6: Fluence-dependent variation of (a) $G_{el}$ and (b) $G_{es}$ considering two different forms for the spin specific heat $C_s$.

## V. Bias field dependence of precessional frequency and Gilbert damping

The bias magnetic field-dependent variation of precessional dynamics in cobalt, nickel and permalloy is shown in Fig. S7 - Fig. S9 for a pump fluence of 4.8 mJ/cm$^2$. The oscillatory Kerr signal is fitted with Eq. 5 of the main text to extract the precessional frequency $f$ and relaxation time $\tau$. The bias field dependence of the extracted frequency, shown in Fig. S7(b) for cobalt, is fitted to the Kittel formula (Eq. 6) to extract the effective saturation magnetization $M_{eff}$. Once $M_{eff}$ have been derived, the Gilbert damping factor $\alpha$ is calculated using Eq. 7 of the main text. The extracted $\alpha$ does not show any systematic variation with bias field.

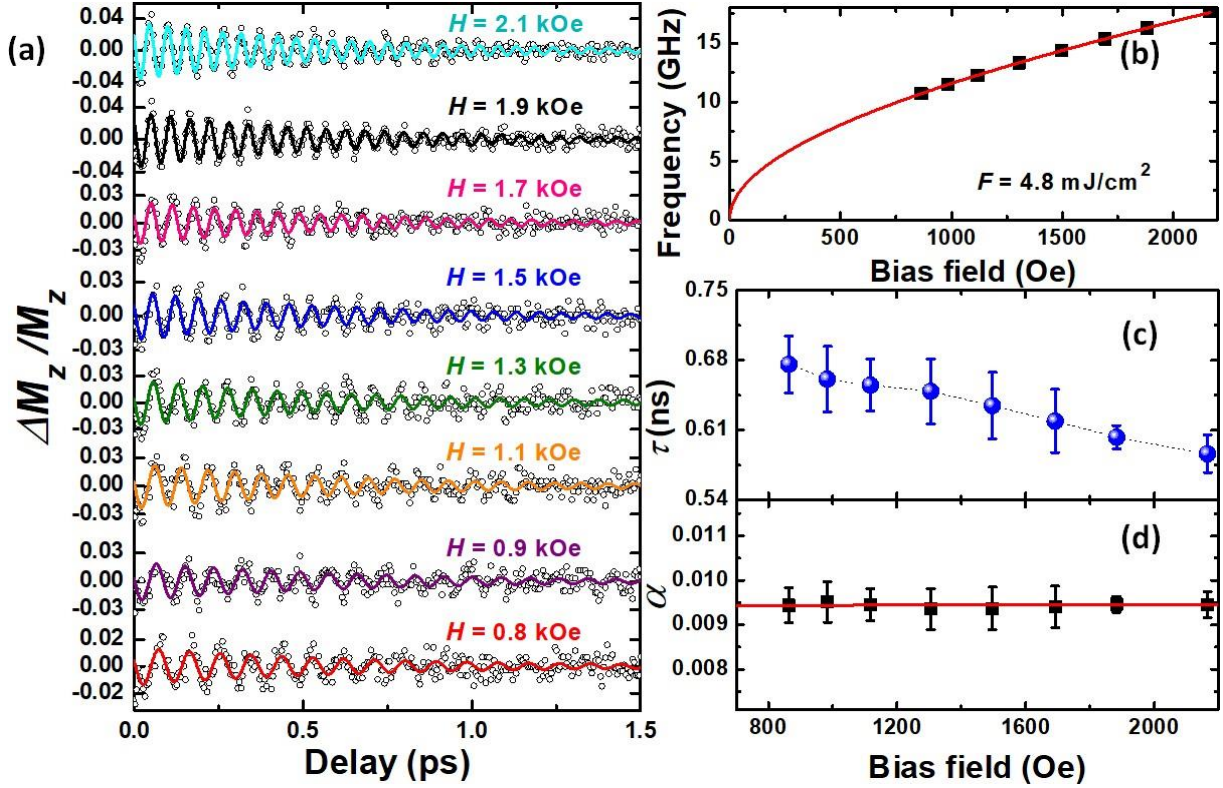

Figure S7: (a) Bias field-dependent variation of precessional dynamics for the cobalt thin film. Symbols are experimental data points and the solid lines are fits to Eq. 5 of the main text. (b) Variation of $f$ with $H$. The solid red line is a fit to the Kittel formula. (c) Variation of $\tau$ with $H$. (d) Variation of $\alpha$ with $H$. The solid line is a guide to the eye.

## VI. Pump fluence dependence of precessional frequency and Gilbert damping

The variation of precessional dynamics in cobalt, nickel and permalloy with the laser pump fluence is shown in Fig. S10 for the cobalt and in Fig. S11 for the permalloy thin film at a saturating field of 2 kOe. The oscillatory Kerr signal is fitted with Eq. 5 of the main text to extract the precessional frequency $f$ and relaxation time $\tau$. With increasing pump fluence, the precessional frequency decreases and the magnetic damping is enhanced. A direct correlation between the demagnetization time $\tau_M$ and $\alpha$ is observed.

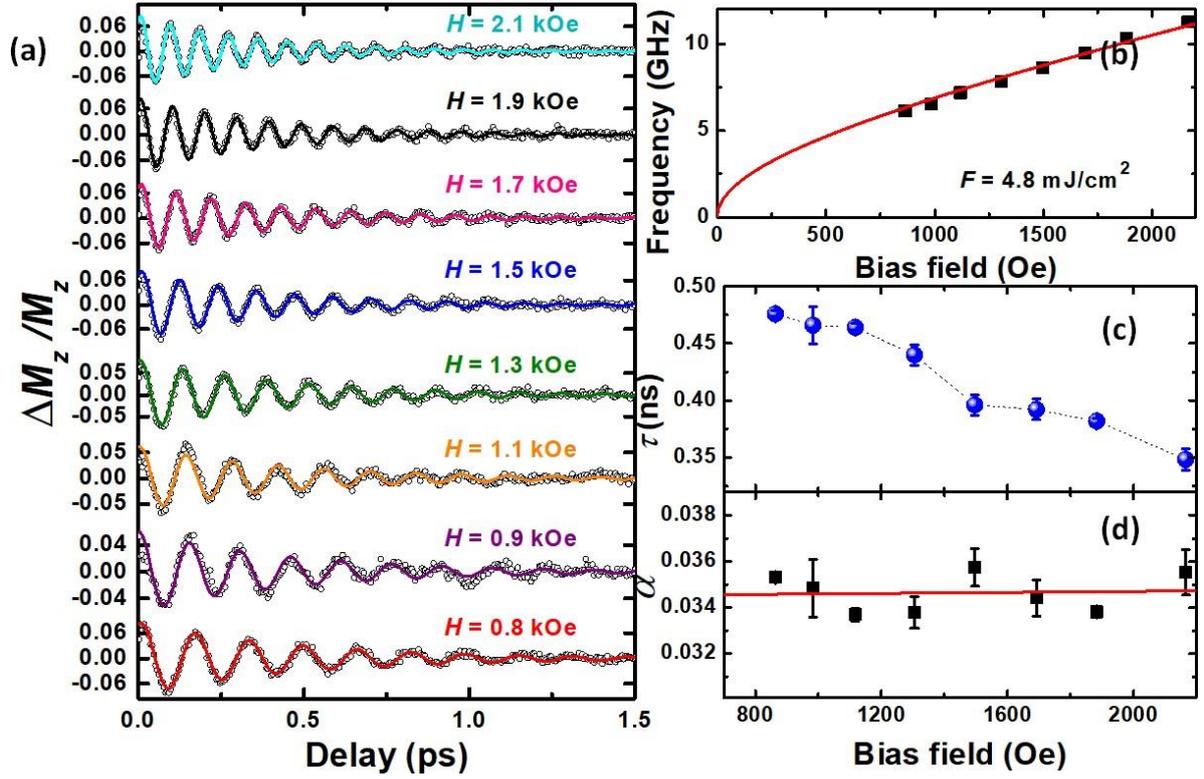

Figure S8: (a) Bias field-dependent variation of precessional dynamics for the nickel thin film. Symbols are experimental data points and the solid lines are fits to Eq. 5 of the main text. (b) Variation of *f* with *H*. The solid red line is a fit to the Kittel formula. (c) Variation of $\tau$ with *H*. (d) Variation of $\alpha$ with *H*. The solid line is a guide to the eye.

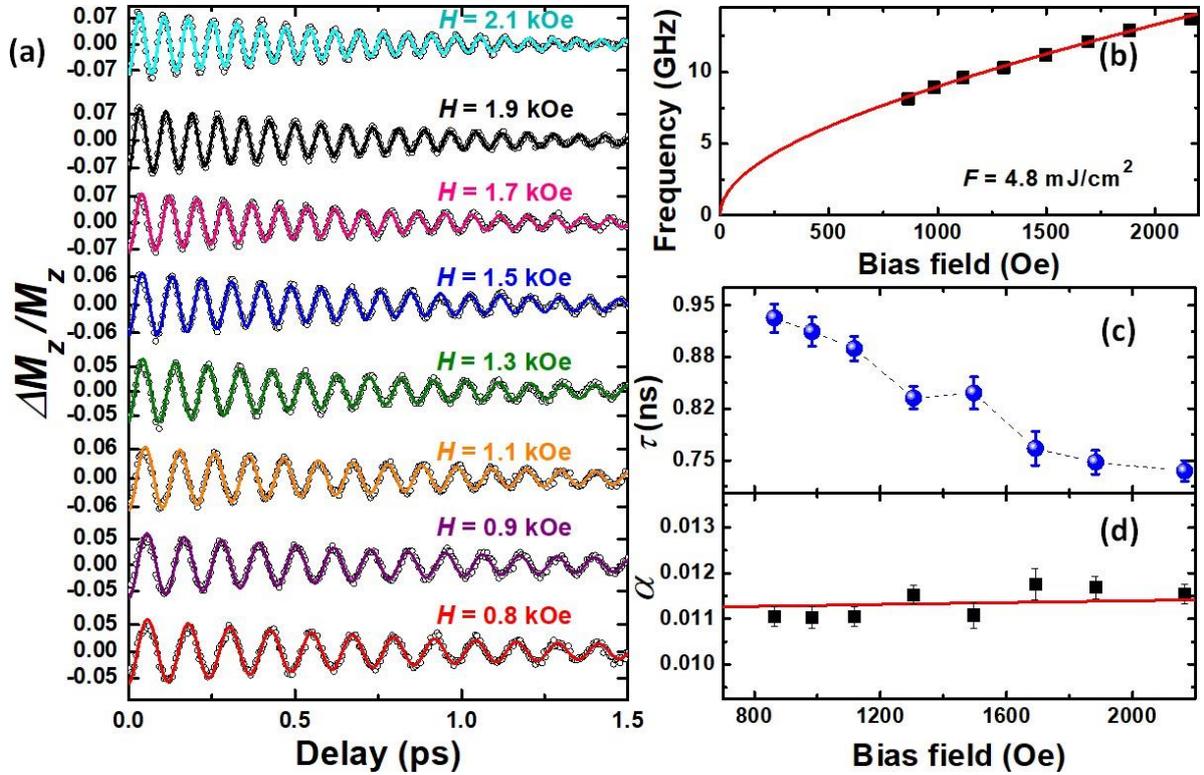

Figure S9: (a) Bias field-dependent variation of precessional dynamics for the permalloy thin film. Symbols are experimental data points and the solid lines are fits to Eq. 5 of the main text. (b) Variation of *f* with *H*. The solid red line is a fit to the Kittel formula. (c) Variation of $\tau$ with *H*. (d) Variation of $\alpha$ with *H*. The solid line is a guide to the eye.

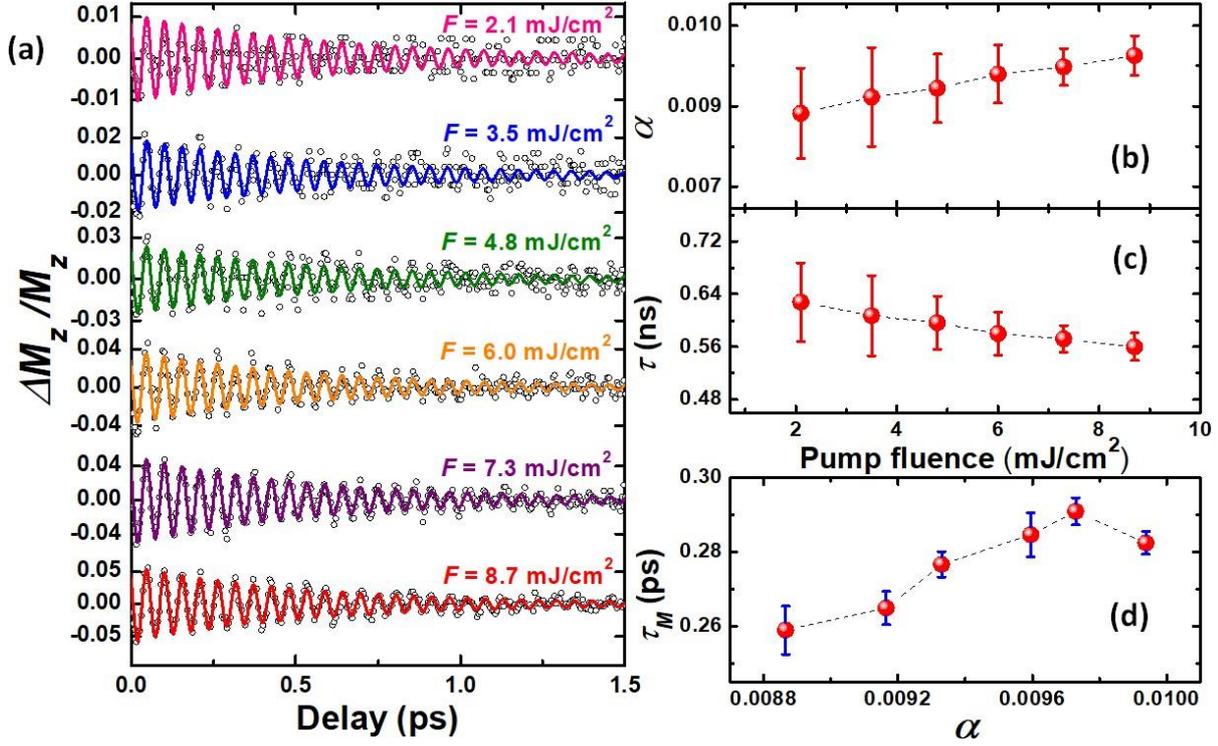

Figure S10: (a) Fluence-dependent variation of precessional dynamics for the cobalt thin film. Symbols are experimental data points and the solid lines are fits to Eq. 5 of the main text. (b) Variation of $f$ with pump fluence. (c) Variation of $\alpha$ with pump fluence. (d) Direct correlation between $\tau_M$ and $\alpha$.

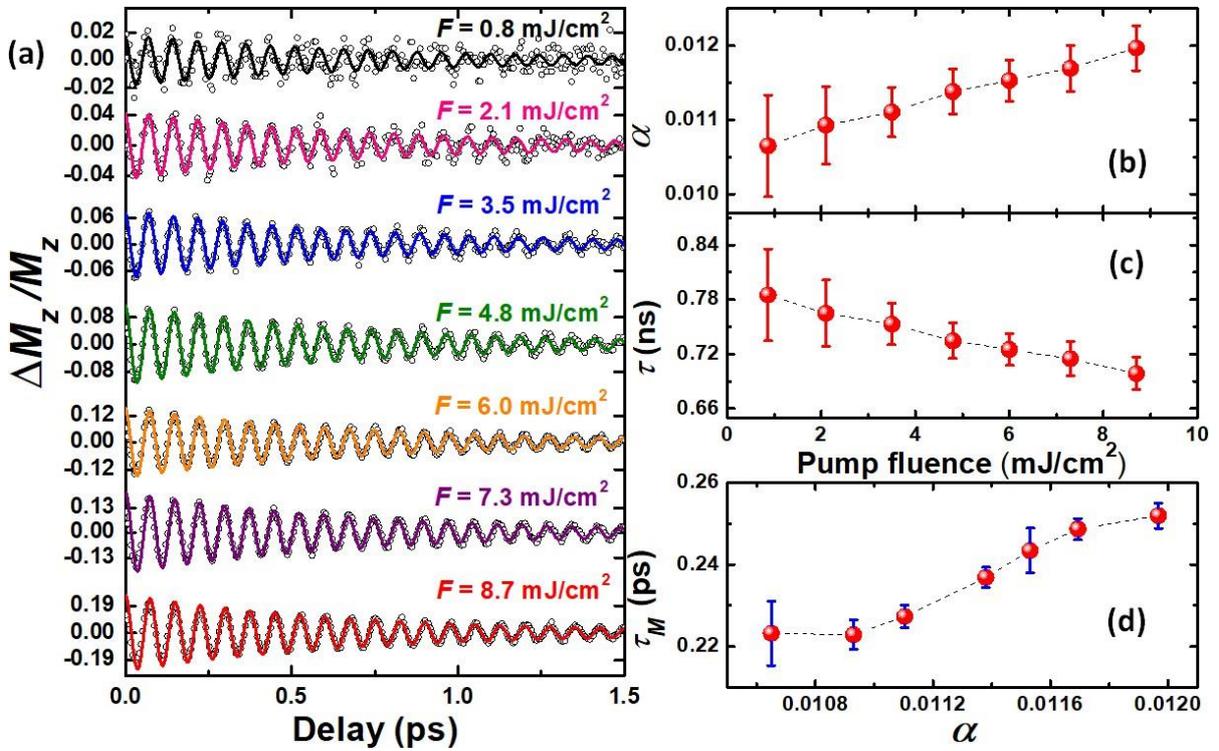

Figure S11: (a) Fluence-dependent variation of precessional dynamics for the permalloy thin film. Symbols are experimental data points and the solid lines are fits to Eq. 5 of the main text. (b) Variation of $f$ with pump fluence. (c) Variation of $\alpha$ with pump fluence. (d) Direct correlation between $\tau_M$ and $\alpha$.